\documentclass[twocolumn]{aastex62}

\usepackage{graphicx}	
\graphicspath{{}}
\usepackage{subfigure}

\usepackage{amsmath}	
\usepackage{amssymb}	

\DeclareRobustCommand{\ion}[2]{%
\relax\ifmmode
\ifx\testbx\f@series
{\mathbf{#1\,\mathsc{#2}}}\else
{\mathrm{#1\,\mathsc{#2}}}\fi
\else\textup{#1\,{\mdseries\textsc{#2}}}%
\fi}

\received{November 23, 2018}
\revised{March 25, 2019}
\accepted{April 30, 2019}
\submitjournal{ApJ}

\shorttitle{The Role of Mergers in the Evolution of LERGs}
\shortauthors{Y.A. Gordon et al.}

\begin{document}

\title{THE EFFECT OF MINOR AND MAJOR MERGERS ON THE EVOLUTION OF \\LOW EXCITATION RADIO GALAXIES}

\correspondingauthor{Yjan A. Gordon}
\email{yjan.gordon@umanitoba.ca}

\author[0000-0003-1432-253X]{Yjan A. Gordon}
\affil{Department of Physics and Astronomy, University of Manitoba, 
Winnipeg, MB R3T 2N2, Canada}
\affil{E.A. Milne Centre for Astrophysics, University of Hull, Cottingham Road, 
Kingston-upon-Hull HU6 7RX, U.K.}

\author{Kevin A. Pimbblet}
\affil{E.A. Milne Centre for Astrophysics, University of Hull, Cottingham Road, 
Kingston-upon-Hull HU6 7RX, U.K.}

\author{Sugata Kaviraj}
\affiliation{Centre for Astrophysics Research, School of Physics, Astronomy and Mathematics, 
University of Hertfordshire, \\College Lane, Hatfield AL10 9AB, U.K.}

\author{Matt S. Owers} 
\affiliation{Department of Physics and Astronomy, Macquarie University, Sydney, NSW 2109, Australia}
\affiliation{Astronomy, Astrophysics and Astrophotonics Research Centre, Macquarie University, Sydney, Australia}

\author{Christopher P. O'Dea} 
\affil{Department of Physics and Astronomy, University of Manitoba, 
Winnipeg, MB R3T 2N2, Canada}
\affil{School of Physics and Astronomy, Rochester Institute of Technology, 
84 Lomb Memorial Drive, Rochester, NY 14623, USA}

\author{Mike Walmsley}
\affil{Department of Physics, University of Oxford, Denys Wilkinson Building, 
Keble Road, Oxford OX1 3RH, U.K.}

\author{Stefi A. Baum}
\affil{Department of Physics and Astronomy, University of Manitoba, 
Winnipeg, MB R3T 2N2, Canada}
\affil{School of Physics and Astronomy, Rochester Institute of Technology, 
84 Lomb Memorial Drive, Rochester, NY 14623, USA}

\author{Jacob P. Crossett}
\affil{School of Physics and Astronomy, University of Birmingham, 
Edgbaston, Birmingham B15 2TT, U.K.}

\author[0000-0001-9557-5648]{Amelia Fraser-McKelvie} 
\affil{School of Physics and Astronomy, University of Nottingham, 
University Park, Nottingham NG7 2RD, U.K.}

\author{Chris J. Lintott} 
\affil{Department of Physics, University of Oxford, Denys Wilkinson Building, 
Keble Road, Oxford OX1 3RH, U.K.}

\author{Jonathon C.S. Pierce}
\affil{Department of Physics and Astronomy, University of Sheffield, Sheffield S3 7RH, U.K.}



\begin{abstract}

We use deep, $\mu_{r} \lesssim 28\,\text{mag}\,\text{arcsec}^{-2}$, $r$-band imaging from the Dark Energy Camera Legacy Survey (DECaLS) to search for past, or ongoing, merger activity in a sample of $282$ Low Excitation Radio Galaxies (LERGs) at $z<0.07$.
Our principle aim is to assess the the role of mergers in the evolution of LERGs.
Exploiting the imaging depth, we classify tidal remnants around galaxies as both minor and major morphological disturbances for our LERG sample and $1,622$ control galaxies matched in redshift, stellar mass, and environment.
In groups and in the field, the LERG minor merger fraction is consistent with the control population.
In galaxy clusters, $8.8 \pm 2.9\,$\% of LERGs show evidence of recent minor mergers in contrast to $23.0\pm 2.0\,$ \% of controls.
This $\sim 4 \sigma $ deficit of minor mergers in cluster LERGs suggests these events may inhibit this type of nuclear activity for galaxies within the cluster environment.
We observe a $> 4\sigma$ excess of major mergers in the LERGs with $M_{*} \lesssim 10^{11}\,\text{M}_{\odot}$, with $10 \pm 1.5\,$\% of these AGN involved in such large-scale interactions compared to $3.2 \pm 0.4\,$\% of control galaxies.
This excess of major mergers in LERGs decreases with increasing stellar mass, vanishing by $M_{*} > 10^{11.3}\,\text{M}_{\odot}$.
These observations show that minor mergers do not fuel LERGs, and are consistent with typical LERGs being powered by accretion of matter from their halo.
Where LERGs are associated with major mergers, these objects may evolve into more efficiently accreting active galactic nuclei as the merger progresses and more gas falls on to the central engine.

\end{abstract} 



\section{Introduction}
\label{intro}
The accretion of matter onto the central supermassive black hole within galaxies powers active galactic nuclei \citep[AGN, e.g.,][]{Salpeter1964, Kaviraj2017}.
When the accretion rate of matter, $\dot M$,  is greater than $\sim 1\,$\% of the Eddington accretion rate, $\dot M_{\text{Edd}}$, the accretion mode can be described as radiatively efficient \citep[e.g.,][]{Jackson1997, Hardcastle2007}.
Such radiatively efficient accretion modes allow for the formation of an optically thick accretion disk that radiates high energy photons \citep[i.e. ultraviolet with upscattering to x-ray, e.g.,][]{Baum1995, Heckman2005}.
These high energy photons act to ionise the local interstellar medium, producing the narrow excitation lines seen in, for example, optically selected AGN \citep[e.g.,][]{Buttiglione2010}.

In addition to the accretion of matter on to the nuclear black hole, in some AGN energy can be further extracted from the central engine by the production of a relativistic jet \citep{Blandford1977}.
Non-thermal radiation from the jet, and its interaction with the surrounding medium, produces relatively bright radio emission in such AGN.
The bulk of these radio-loud AGN (RLAGN) at low redshift show no evidence of high-excitation lines resulting from an optically thick nuclear accretion mechanism \citep{Best2012}.
These Low Excitation Radio Galaxies (LERGs) are thus thought to not possess the same type of accretion disk associated with radiatively efficient accretion modes, and instead are fuelled by an optically thin, advection dominated accretion flow \citep{Fabian1995, Narayan1995}.
In order to explain this radiatively inefficient accretion whilst still producing the AGN jet, very low Eddington scaled accretion rates, i.e. $\dot M \ll 0.01\dot M_{\text{Edd}}$ are invoked \citep[e.g.,][]{Baum1992, Baum1995, Tadhunter1998, Hardcastle2006, Allen2006, Evans2011, Mingo2014}.

The weakly accreting nature of LERGs is indicative of a poorer fuel supply than is available to more efficiently accreting AGN \citep[e.g.,][]{Best2012, Ellison2015}.
Furthermore, LERGs are usually passive in terms of star formation, and associated with red, massive, early-type galaxies \citep[e.g.,][]{Heckman1986, Best2005a, Kauffmann2008, Lofthouse2018}.
This deficiency of star formation adds further weight to the argument that LERGs lack a ready supply of cold gas.
The environments that host LERGs are frequently observed to be over-dense, with LERGs often being brightest cluster galaxies \citep[e.g.,][]{Hill1991, Zirbel1997, Best2007, RamosAlmeida2013, Ching2017}.
In such dense environments, the hot intra-cluster medium (ICM) acts to inhibit the accretion of cold gas by galaxies \citep{Davies2017}.

Limited supplies of cold gas can be accreted onto galaxies in cluster cores via cooling flows \citep[e.g.,][]{ODea1994, ODea2008, Edge2001, Pipino2009, Donahue2011} and hence provide a potential fuel supply for an AGN.
Such cooling flows may be enhanced by thermal instabilities in a dynamic ICM, a process known as chaotic cold accretion \citep[CCA,][]{Gaspari2013, Gaspari2017}.
Additionally, AGN driven outflows may cool as they expand, allowing gas to fall back on to, and drip-feed the central engine \citep[e.g.,][]{Tremblay2016, Voit2017, Tremblay2018}.
Finally, galactic mergers present an obvious mechanism with which to introduce a cold gas fuel reservoir to power the AGN \citep{Sanders1988, Weston2017}, although evidence for this mechanism is mixed \citep[e.g.,][]{Scott2014, Villforth2017}.
Given the expectation for a limited fuel supply in LERGs, then if mergers are involved in their triggering they should either be gas-poor, or else indirect triggers rather than a direct fuel supply.
In this scenario, minor mergers present an attractive trigger mechanism for LERGs \citep[e.g.,][]{Kaviraj2014, Pace2014, Ellison2015, Martin2018}, and might be expected given the excess of satellites observed around LERGs \citep{Pace2014}.
Such low-impact galactic collisions would provide a restricted gas supply that may fall short of initiating the radiatively efficient accretion modes associated with High Excitation Radio Galaxies (HERGs) and AGN selected from non-radio bands.

In this work we aim to test this last hypothesis that mergers, and in particular minor mergers, play a role in the evolution of LERGs.
The low mass ratio involved in a minor merger \citep[$\lesssim$1:4,][]{Lotz2010} results in a limited impact on the morphology of the primary, or recipient, galaxy in the merger.
The morphology of the secondary (hereafter donor) galaxy is totally disrupted as it is absorbed by the recipient galaxy.
In combination these effects can make detecting minor mergers problematic in the relatively shallow imaging obtained by typical wide-field galaxy surveys \citep{Kaviraj2010, Kaviraj2014a}.
Instead, observational evidence for minor mergers presents as subtle low surface brightness (LSB) tidal features (e.g., tails, streams and halo shells), the results of stellar material stripped from the donor galaxy during its infall onto the recipient galaxy \citep[e.g.,][]{RamosAlmeida2012, Kaviraj2014}.

Detecting LSB features requires deeper imaging than is necessary for normal morphological studies of massive galaxies.
A new range of wide-deep imaging surveys such as Stripe 82 of the Sloan Digital Sky Survey \citep[SDSS,][]{York2000, Fliri2016}, the Kilo Degree Survey \citep[KiDS,][]{DeJong2013}, and the Dark Energy Camera Legacy Survey \citep[DECaLS,][]{Blum2016}, are paving the way for large studies of LSB structures such as tidal features.
One would expect that, should mergers be involved in the evolution of a galaxy in to a LERG, then an excess of merger signatures would be observed in LERGs compared to a control sample of galaxies.
Exploiting the latest in deep imaging surveys allows this approach to be extended to minor mergers by comparing the rates of LSB morphological disruption.
This method was used by \citet{Kaviraj2014a} to demonstrate the significant role of minor mergers in fuelling star formation at low redshift, and it is this procedure we employ in this paper.

The layout of this paper is as follows. 
In Section \ref{data}, we detail the datasets used in this paper and describe the LERG and control sample selections.
Section \ref{imaging} describes the process of classifying the DECaLS images. 
We state our results in Section \ref{obs} and discuss the implications of these observations in Section \ref{discussion}. 
Section \ref{sum} is a summary of this work. 
Throughout this work we assume a standard flat $\Lambda$CDM cosmology with $h=0.7$, $H_{0}=100h\,\text{km}\,\text{s}^{-1}\,\text{Mpc}^{-1}$, $\Omega_\text{M} = 0.3$, $\Omega_\Lambda = 0.7$.


\section{Observational Datasets and Sample Selection}
\label{data}
To assess the role of mergers in the evolution of LERGs requires optical spectra and imaging, as well as radio observations.
Optical spectroscopic data is obtained from SDSS data release 7 \citep[DR7,][]{Abazajian2009}, containing $\sim 10^{6}$ spectra of galaxies and quasars across $\sim 8,000\,\text{deg}^{2}$ of the sky.
Optical imaging is obtained from data release 5 (DR 5) of DECaLS, which broadly covers the region $-22^{o} < \delta < +34^{o}, |b|>18^{o}$ at $r \lesssim 24\,$mag.
In the $\sim 5,000\,\text{deg}^{2}$ where the survey footprints overlap, DECaLS provides imaging that is approximately 2 mags deeper than the standard SDSS optical imaging.
Beyond point source detection, it is the enhanced ability of DECaLS over SDSS to detect low surface brightnesses that is important in this work.
In comparison to the standard depth SDSS imaging which can detect surface brightnesses of $\mu_{r} \sim 25\,\text{mag}\,\text{arcsec}^{-2}$ \citep{Driver2016}, DECaLS observes surface brightnesses of $\mu_{r} \lesssim 28\,\text{mag}\,\text{arcsec}^{-2}$ \citep{Hood2018}.

\subsection{LERG Selection}
In addition to the optical data, radio observations are required in order to detect and classify LERGs.
To this end we select our LERGs from the \citet{Best2012} catalogue of 18,286 radio galaxies in SDSS. 
This catalogue is the result of cross matching observations from SDSS, the National Radio Astronomy Observatory (NRAO) Very Large Array (VLA) Sky Survey \citep[NVSS,][]{Condon1998}, and the VLA Faint Images of the Radio Sky at Twenty cm survey \citep[FIRST,][]{Becker1995, Best2005, Best2012}. 
The \citet{Best2012} catalogue segregates radio galaxies in to those where the radio emission is the result of star formation from those where it is the result of an AGN. 
This is based on a combination of  stellar mass, $1.4\,\text{GHz}$ luminosity, H$\alpha$ luminosity, $4,000\,\text{\AA}$ break strength, and emission line diagnostics \citep{Baldwin1981, Kauffmann2003, Best2005, Kauffmann2008, Best2012}.
Radio-loud AGN are then further classified into high- or low-excitation sources based on the host galaxy spectrum. For high-quality spectra with many observed emission lines, the excitation index of \citet{Buttiglione2010} is used for the purpose.
For poorer quality spectra, or those with intrinsically fewer emission lines, a more simple $\text{EW}_{\text{[\ion{O}{iii}]~$\lambda5007$}} > 5\,\text{\AA}$ criteria is used to segregate HERGs and LERGs.
For a full description of the radio galaxy classification used in the construction of their catalogue, the reader is directed to \citet{Best2012}, and \citet{Best2005}.

\begin{figure}
	\centering
	\includegraphics[width=\columnwidth]{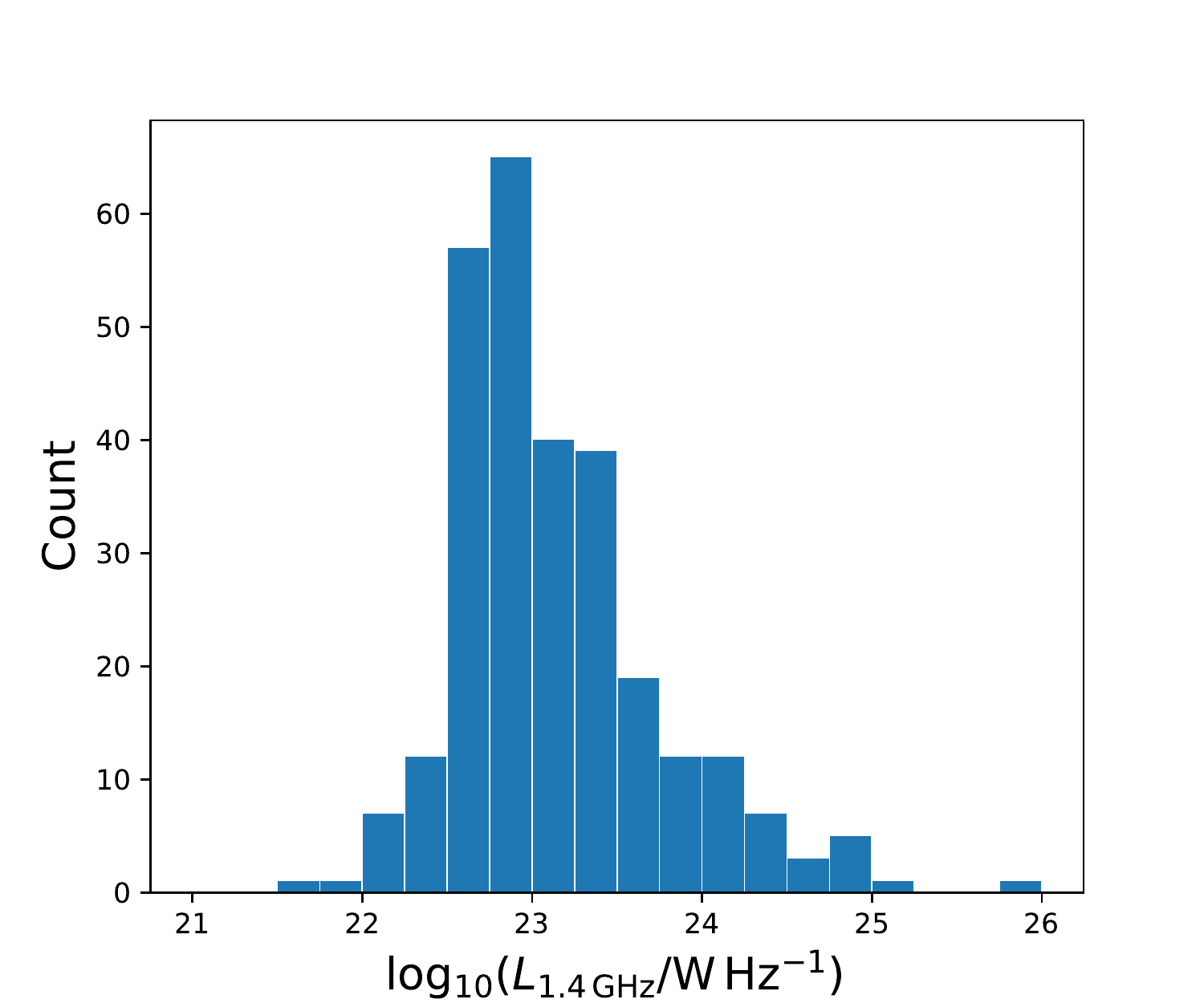}
	\caption{The distribution of $1.4\,\text{GHz}$ luminosities of our LERG sample.}
	\label{rpowerdist}
\end{figure}

The nature of LSB astronomy necessitates that only local-Universe galaxies can be included in our analysis.
For this reason we select LERGs from the \citet{Best2012} catalogue with $z<0.07$ \citep{Kaviraj2014a}.
The LERGs selected cover four orders of magnitude in radio luminosity, $21.7 < (L_{1.4\,\text{GHz}}/\text{W}\,\text{Hz}^{-1}) < 25.8$, with a median luminosity of $10^{23}\,\text{W}\,\text{Hz}^{-1}$ at $1.4\,$GHz. 
The distribution of our LERG radio luminosities is shown in Figure \ref{rpowerdist}. 
Additionally, their distribution in redshift, stellar mass, group/cluster halo mass (where applicable), and colour are shown alongside the control sample (described below) distributions in Figure \ref{LERGcontDists}.

\subsection{The Main Control Sample}
\label{controlDescription}
\subsubsection{Redshift and stellar mass}
In order to assess whether or not there is an excess of tidal remnants around LERGs, a suitable control sample of radio-quiet galaxies must be constructed. To this end we match each LERG to galaxies without a radio-detection on the basis of redshift and stellar mass.
Stellar mass estimates for both the LERGs and control galaxy candidates are taken from the MPA/JHU value added catalogue for SDSS DR7\footnote{\url{https://wwwmpa.mpa-garching.mpg.de/SDSS/DR7/}} \citep{Tremonti2004}.
These stellar masses are calculated adopting the method of \citet{Kauffmann2003a} but using the SDSS $ugriz$ photometry of the source rather than spectral indicies\footnote{A comparison to the SDSS stellar masses obtained by \citet{Kauffmann2003a}, who used spectral features rather than broad band photometry, can be found at \url{https://wwwmpa.mpa-garching.mpg.de/SDSS/DR7/mass_comp.html}}.
We require that control galaxy candidates have:
\begin{itemize}
\item $z_{\text{control}} = z_{\text{LERG}}\pm <0.01$
\item $M_{*, \text{control}} = M_{*, \text{LERG}} \pm < \sigma_{M_{*}}$,
\end{itemize}
Where $\sigma_{M_{*}}$ is obtained from the 16th and 84th percentiles of the probability distribution of the stellar mass estimates.

Photometrically derived stellar mass estimates can be biased in AGN due to the excess short-wavelength radiation from the accretion disk \citep[this is especially true in unobscured AGN,][]{Gordon2017}, and indeed powerful radio galaxies often exhibit an excess of ultraviolet light \citep{Tadhunter2002}.
However, given that LERGs are inefficiently accreting AGN, and thus do not have an optically thick, luminous accretion disk, contamination of the broadband optical photometry from the AGN should be insignificant.
Furthermore, \cite{Kauffmann2008} demonstrated that for the \citet{Best2005} sample of radio galaxies in SDSS, which covers a radio power range inclusive of our LERG sample's $L_{1.4\,\text{GHz}}$ distribution, such a UV excess was consistent with originating from a young stellar population rather than from AGN contamination.
Thus, stellar mass estimates for our LERG population should be as reliable as the stellar mass estimates for galaxies not hosting an AGN.
Moreover, obtaining stellar masses from the MPA/JHU catalogue for our LERGs is consistent with previous works \citep{Smolcic2009, Best2012}.

\subsubsection{Large-scale structure}
Beyond controlling for just stellar mass and redshift, we require that each LERG is matched with control galaxies in the same type of large-scale environment.
To determine this we use the \citet{Yang2007} SDSS group catalogue for DR7. 
For those galaxies within this catalogue we use the halo mass, $M_{180}$, of the structure the galaxy is located in to characterise its environment, considering halos of $M_{180} > 10^{12.5}\,\text{M}_{\odot}$ to be a group, and halos more massive than $10^{14}\,\text{M}_{\odot}$ to be a cluster \citep{Gordon2018, Barsanti2017, Lofthouse2018}.
\citet{Yang2007} observe that the error in obtained halo masses for their catalogue vary between 0.2 and 0.35 dex, and remain above $0.25\,$dex for the mass range $12.2 \lesssim \log_{10}(M_{180}/\text{M}_{\odot}) \lesssim 14.6$ (see their Fig. 7).
By matching halos with $\Delta M_{180} < 0.25\,$dex, we are thus matching them to other halos of reliable comparable mass.
Galaxies in halos of $M_{180} < 10^{12.5}\,\text{M}_{\odot}$ are treated as field galaxies and are matched only with other field galaxies.

\begin{figure}
	\centering
	\includegraphics[width=\columnwidth]{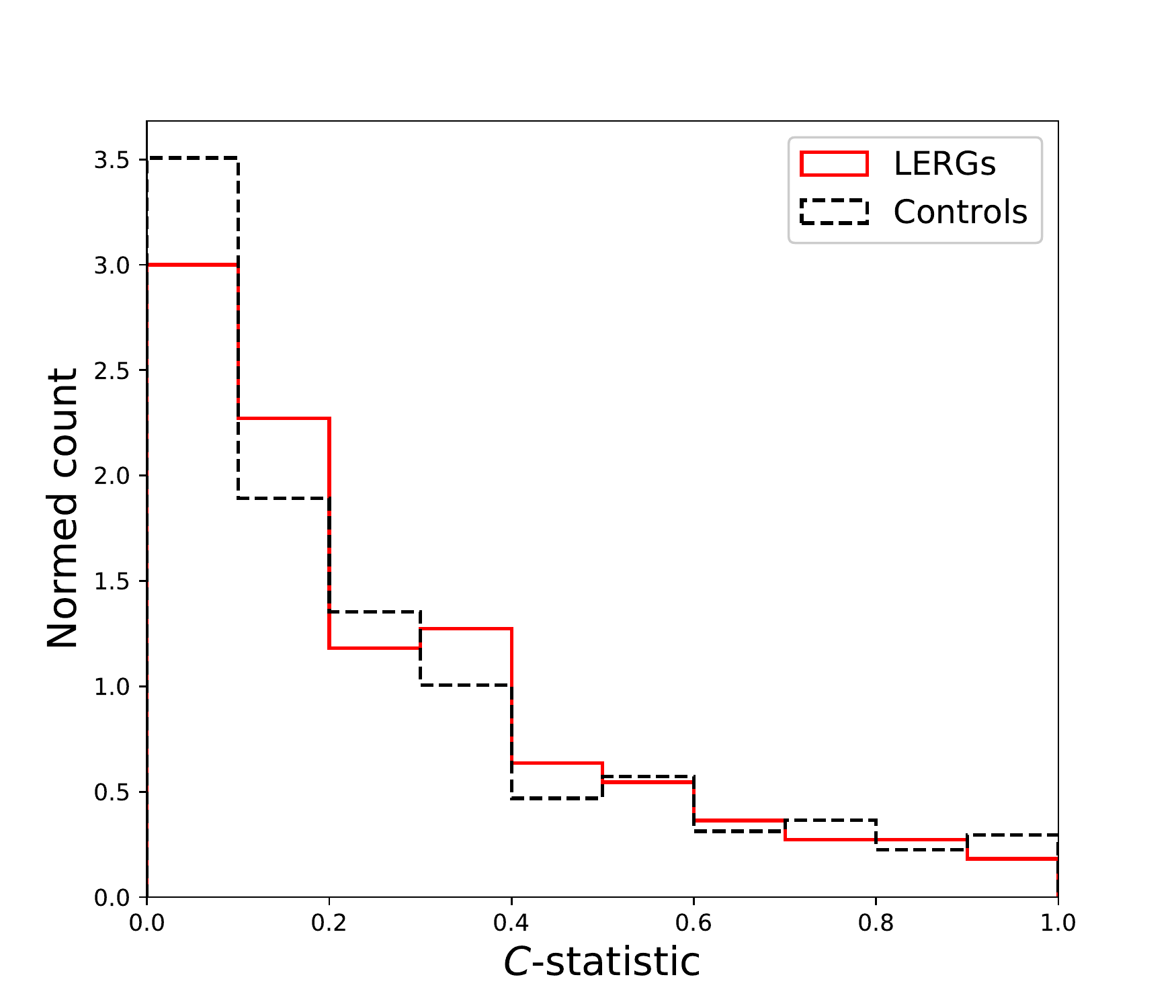}
\caption{The distribution of calculated $C$-statistic \citep{Smith2004} values for our selected galaxies assigned membership to large-scale structures, where a lower value is associated with increased likelihood of membership.
The red solid line shows the $C$ distribution for the LERG population, and the black dashed line for the control population.}
\label{cstatdists}
\end{figure}

Given the known tendency for LERGs to be found within dense environments \citep[e.g.][]{Ching2017}, it may be possible that our LERGs observed to lie within large scale structures may be more accurately assigned to groups and clusters than their controls. 
That is to say, the probability of group or cluster membership for a galaxy found by the group finding method of \citet{Yang2007} may be higher if the galaxy is a LERG.
To determine if such an effect exists within our selected LERG and control populations, we calculate the $C$-statistic of \citet{Smith2004} for each of our selected galaxies.
The $C$-statistic is a measure of likelihood of a galaxy to be associated with a particular structure and is given by
\begin{equation}
C = \frac{(cz_{\text{galaxy}} - cz_{\text{cluster}})^{2}}{\sigma^{2}} - 4\log_{10}\Bigg(1 - \frac{R}{R_{\text{cluster}}}\Bigg),
\end{equation}\
where, $z_{\text{galaxy}}$ is the galaxy redshift, $z_{\text{cluster}}$ is the median cluster (or group) redshift,  $\sigma$ is the cluster velocity dispersion, $R$ is the projected separation of the galaxy from the cluster centre, and $R_{\text{cluster}}$ is the cluster radius.
In Figure \ref{cstatdists} the distributions of this statistic for both our LERG and control samples associated with large scale structures are shown to be similar for both populations.
Additionally, these distributions fail to separated by Kolmogorov-Smirnov (KS) testing, providing a $p$-value of 0.4, demonstrating that there is no significant difference in the accuracy of group and cluster membership assignment between our LERG and control populations.

\begin{figure*}
	\centering
	\subfigure[]{\includegraphics[width=\columnwidth]{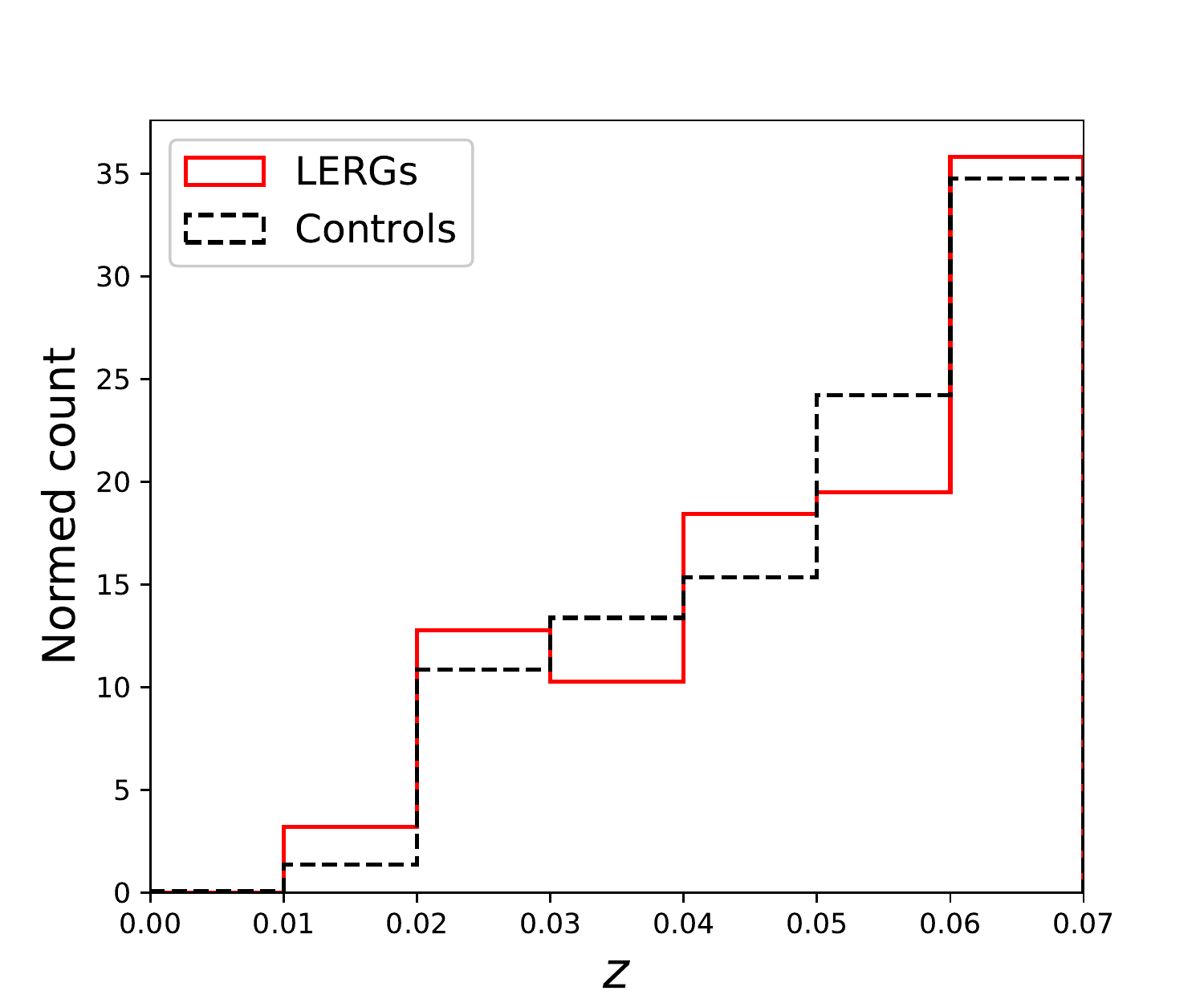}}
	\subfigure[]{\includegraphics[width=\columnwidth]{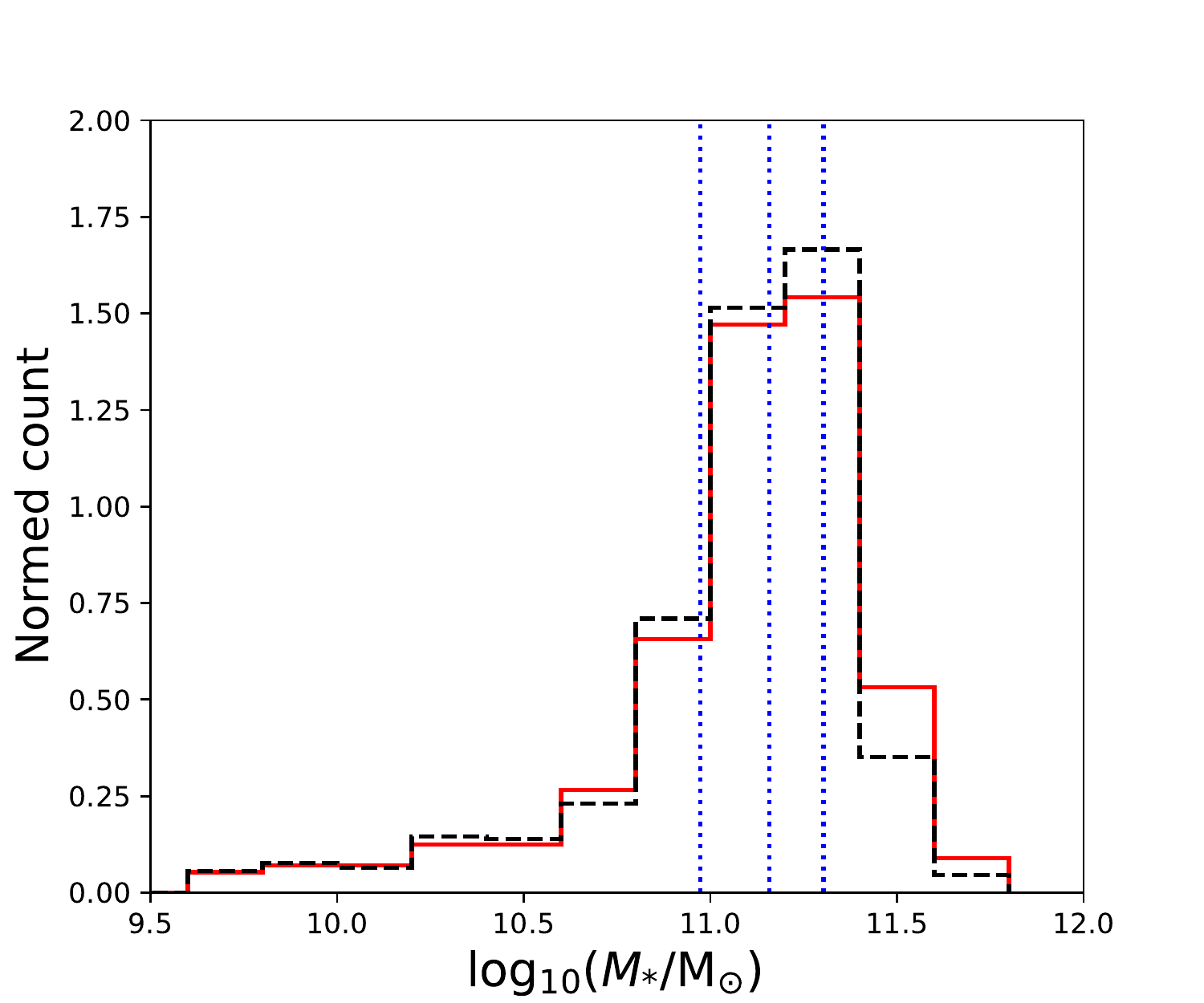}}
	\subfigure[]{\includegraphics[width=\columnwidth]{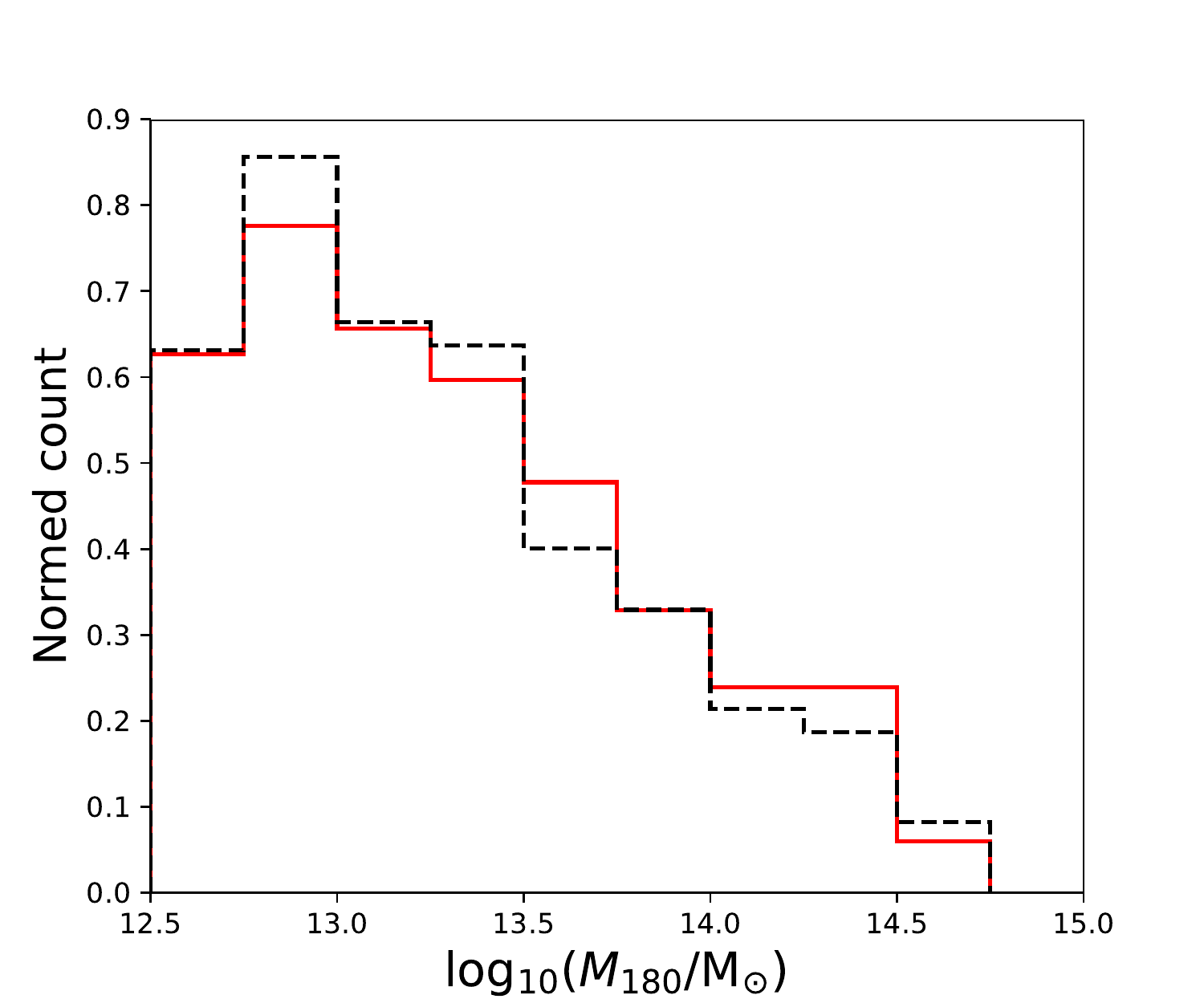}}
	\subfigure[]{\includegraphics[width=\columnwidth]{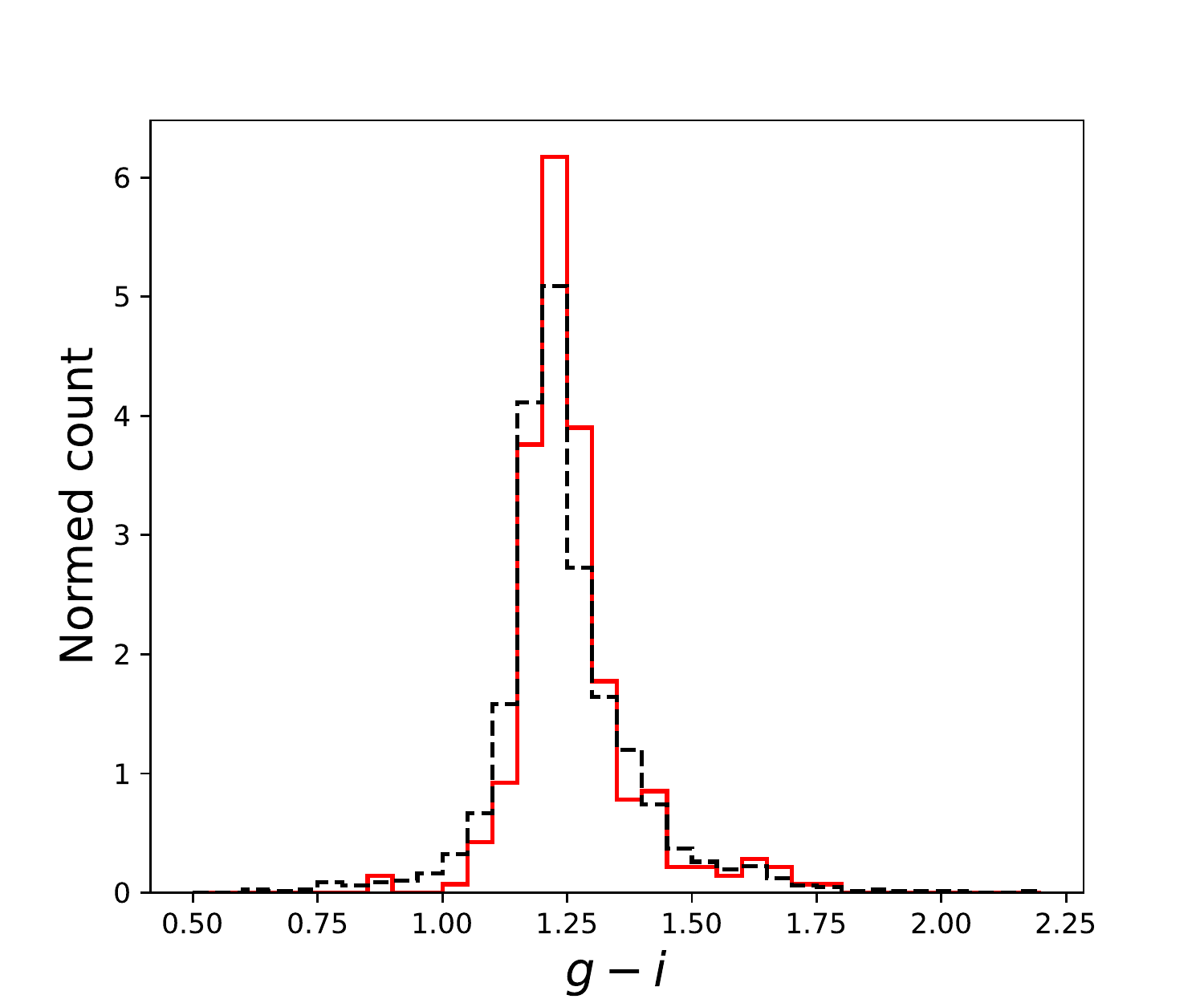}}
\caption{The similar distributions of the redshift (panel a), stellar mass (panel b), group halo masses (panel c), and the $g-i$ colours (panel d) of the LERG and control samples. The red solid line represents LERGs and the black dashed line represents the control population. The blue dotted lines shown in panel b highlight the 25th, 50th, and 75th percentiles of the LERG stellar mass distribution, used for mass subsetting in our analysis.}
\label{LERGcontDists}
\end{figure*}

For each LERG we select up to six (where possible) control galaxies satisfying these criteria, prioritised by closest match in stellar mass.
Where fewer than six control galaxies for a LERG are found, as many controls as possible that satisfy the matching criteria are selected \footnote{The weighting of controls in such cases is discussed in-depth in Section \ref{ControlWeights}.}.
This results in a parent sample of $1,648$ control galaxies matched to $284$ LERGs.
Whilst we do not control based on galaxy colour, we note that the $g-i$ colour distribution of our control sample is similar to that of our LERGs.
This, alongside the LERG and control sample distributions of redshift, stellar mass and, for non-field galaxies, halo mass are shown in Figure \ref{LERGcontDists}.

\subsection{Controlling for Galaxy Morphology}
\label{galzoo}
Ideally the control sample should also be matched to the LERG population on morphology.
To this end, we obtain morphological information for our selected LERGs and controls as part of the ongoing Galaxy Zoo project (\citealp{Lintott2008, Willett2013}; Walmsley et al. in prep).
The current iteration of Galaxy Zoo\footnote{\url{https://www.zooniverse.org/projects/zookeeper/galaxy-zoo}} has the advantage of using colour images from DECaLS rather than the SDSS images used by previous versions of the project.
Whilst Galaxy Zoo provides data on various subtle morphological parameters \citep[e.g., number of spiral arms,][]{Willett2013}, for this work we wish only to know if a galaxy is early- or late-type.

To classify our galaxies as early- or late-type, we require at least $95\,$ \% confidence in voting for one particular morphology over another, where the confidence limits assume binomial errors \citep{Cameron2011} and are calculated from the raw number of votes for each answer.
The initial question in the current Galaxy Zoo workflow (to be described in full in Walmsley et al, in prep, and based on \citealp{Willett2013}), is concerning the broad morphology of the galaxy, i.e. is it \textit{smooth and rounded} (early-type) or \textit{disk or featured} (late-type).
To classify our galaxies as either early- or late-type we require that the \text{$95\,$ \%} confidence intervals for these two answers do not overlap.
Additionally we require that, for the favoured classification, the voting indicates a majority verdict (i.e. $>50\,$\% of the total votes) at greater than $80\,$\% confidence.

Using this method, we obtain reliable morphologies for 216 of our LERGs and 1,138 of the control galaxies. 
Of our LERGs with reliable classifications $91\,$\% are classified as early-type, compared to $67\,$ \% of the control sample.
Ensuring the control galaxy morphology is the same as the LERG morphology (in addition to satisfying the other control parameters described in Section \ref{controlDescription}) results in a morphologically-matched subset of 191 LERGs and 657 controls.
Whilst our primary analysis is conducted on the overall LERG and control samples, this morphologically-matched subset provides a benchmark to compare our results with and thus ensure that our findings are not biased by galaxy morphology.


\section{Analysis of DECaLS Imaging}
\label{imaging}

\begin{figure*}
	\centering
	\subfigure{\includegraphics[width=0.15\textwidth]{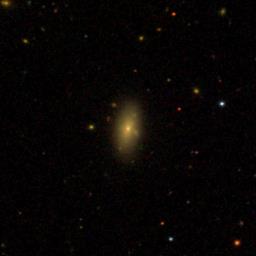}}
	\subfigure{\includegraphics[width=0.15\textwidth]{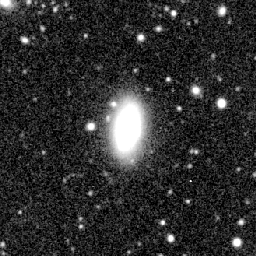}}
	\hspace{1mm}
	\subfigure{\includegraphics[width=0.15\textwidth]{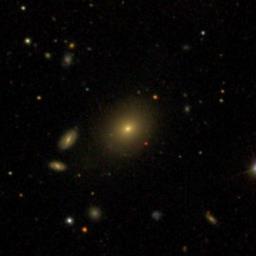}}
	\subfigure{\includegraphics[width=0.15\textwidth]{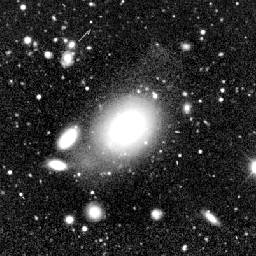}}
	\hspace{1mm}
	\subfigure{\includegraphics[width=0.15\textwidth]{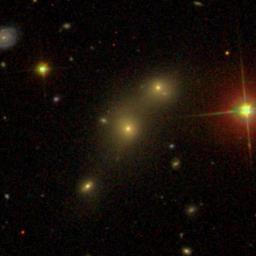}}
	\subfigure{\includegraphics[width=0.15\textwidth]{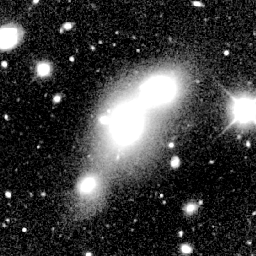}}
	
	\vspace{-3mm}
	
	\subfigure{\includegraphics[width=0.15\textwidth]{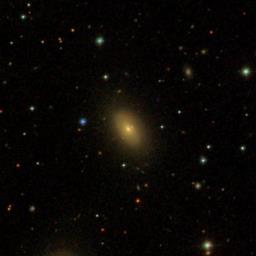}}
	\subfigure{\includegraphics[width=0.15\textwidth]{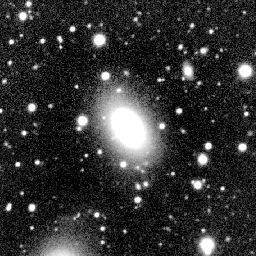}}
	\hspace{1mm}
	\subfigure{\includegraphics[width=0.15\textwidth]{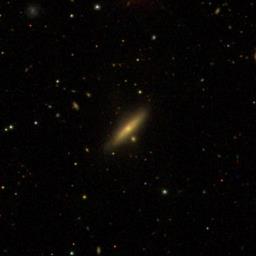}}
	\subfigure{\includegraphics[width=0.15\textwidth]{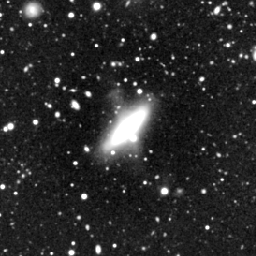}}
	\hspace{1mm}
	\subfigure{\includegraphics[width=0.15\textwidth]{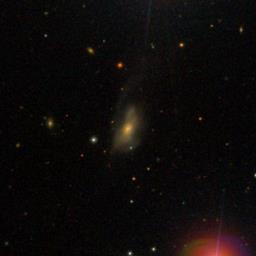}}
	\subfigure{\includegraphics[width=0.15\textwidth]{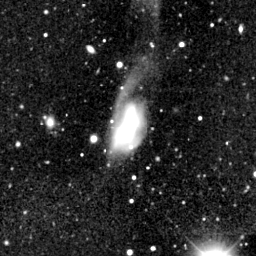}}
	
	\vspace{-3mm}
	
	\subfigure{\includegraphics[width=0.15\textwidth]{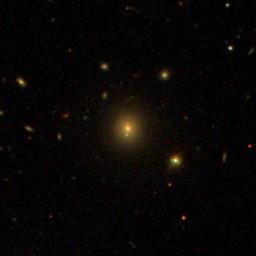}}
	\subfigure{\includegraphics[width=0.15\textwidth]{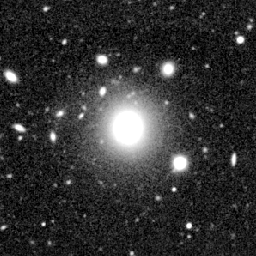}}
	\hspace{1mm}
	\subfigure{\includegraphics[width=0.15\textwidth]{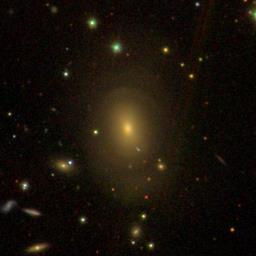}}
	\subfigure{\includegraphics[width=0.15\textwidth]{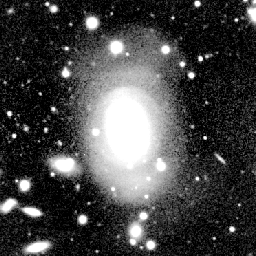}}
	\hspace{1mm}
	\subfigure{\includegraphics[width=0.15\textwidth]{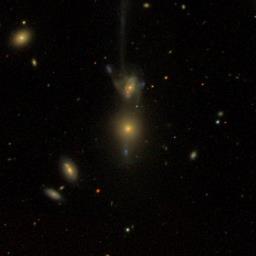}}
	\subfigure{\includegraphics[width=0.15\textwidth]{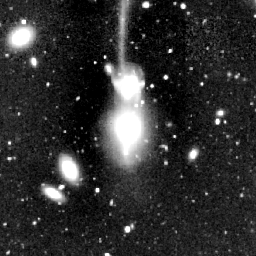}}	
	
\caption{Example images from SDSS and DECaLS of galaxies of different classifications from our analysis, three examples of each classification are given - one per row. Split into 3 columns, each 2 panels wide, the left panel in a column shows an SDSS standard-depth colour image of the galaxy, whilst the right panel in a column shows the deeper DECaLS DR5 $r$-band image used in this work. The left-most column shows galaxies classified as having no morphological disturbance, the middle column shows minor morphological disturbances, and the right-most column shows major morphological disturbances. Note how for the minor, and even the major, disturbances the DECaLS images readily show LSB features that are barely noticeable or indeed totally absent in the SDSS imaging.}
\label{egClass}
\end{figure*}

For each of the LERGs and control galaxies, a DECaLS DR5 \text{$r$-band} cutout, measuring $200 \times 200\,$kpc at the redshift of the target galaxy, is obtained. 
There are some small regions in DECaLS DR5 where the imaging in all three bands is incomplete, resulting in a small fraction ($\sim1\,$\%) of our data selection not having DECaLS DR5 $r$-band imaging.
The images were processed using an arcsinh stretch, allowing the fainter details to be seen whilst preserving the brighter features better than is possible using a logarithmic scaling \citep{Lupton2004}. 
The resultant `postage stamp' image is then uploaded to the Zooniverse project builder\footnote{\url{https://www.zooniverse.org/lab}}, where the images are presented in a random order for blind classification.
We show examples of these in Figure \ref{egClass} alongside SDSS colour images for comparison.
Each image presented had four possible options to vote for, of which only one could be selected:
\begin{itemize}
\item`\textit{No disturbance}' required that there be no obvious asymmetries to the low- or high-surface brightness features, no apparent tidal tails or shocks (e.g., column one of Figure \ref{egClass}).\\
\item `\textit{Minor disturbance}' was selected should an image have features affecting the LSB morphology, e.g., halo shells or faint tidal streams (see column two of Figure \ref{egClass}).\\
\item `\textit{Major disturbance}' was dependent on there being clear disruption to the high surface brightness morphology of the galaxy, frequently with second similar size galaxy involved (as shown in column three of Figure \ref{egClass}).\\
\item `\textit{Bad data}' was included as an option should the image quality prevent classification (e.g. due to bad stitching or artefacts).\\
\end{itemize}

The classification of the images is a subjective task for which there may be variance between individual voters.
However, as each person's voting was conducted independently, individual analyses should be self consistent. 
In total seven volunteers from the coauthorship were involved in the classification process, providing $10,796$ votes for a total of $1,904$ images, with each image having been classified by between five and seven of these classifiers. 


\section{Results}
\label{obs}
\subsection{The Prevalence of Tidal Features in LERGs}
Images of $282$ LERGs and $1,622$ control galaxies were analysed.
The votes from each classifier were stacked and the fraction of each of the possible classifications calculated.
Uncertainties are estimated by bootstrapping the result $10,000$ times, and using the 16th and 84th percentiles of the resultant distribution as the lower and upper errors \footnote{Using $P_{16}$ and $P_{84}$ as the error estimates here, rather than the standard deviation, allows for potential variance in the symmetry of the bootstrapped distribution when calculating the $68\,\%$ confidence interval.}.
We find that the overall merger fractions of LERGs and controls are broadly consistent, with \text{$28.7 \pm1.1\,$\%} of LERGs and $27.3 \pm 0.5\,$\% of controls observed to have tidal features.
In the LERG sample, major tidal features are found in $7.2 \pm 0.6\,$\% of the population, while $21.4_{-1.0}^{+1.1}\,$\% of LERGs are observed to have minor morphological disturbances.
For the control sample, $5.0 \pm 0.2\,$\% have major morphological disturbances, and $22.3 \pm 0.5\,$\% have minor tidal remnants.
Whilst minor disruption is seen in the LSB morphologies of LERGs and controls at similar rates, there is an excess of major tidal features observed in the LERG population at $>3\sigma$ confidence. 
The fractions of LERGs and controls with major and minor tidal features are shown in Table \ref{Tfrac}.

Our morphologically-matched subsample shows similar results to the analysis conducted without controlling for morphology. 
For these LERGs we report that $5.3 \pm 0.7\,$\% and $17.4 \pm 1.1\,$\% are associated with major and minor tidal features respectively. 
Their control sample shows $3.4 \pm 0.3\,$\% observed to have major tidal remnants, and $20.8 \pm 0.7\,$\% with minor morphological disruption.
This suggests, with $\sim 2.5\sigma$ confidence, that the tendency for LERGs to be more likely than controls to have substantial tidal features is not dependent on the galaxy morphology.
We attribute the systematically lower fractions of galaxies with tidal features in this morphologically-matched subsample to the need for a reliable morphological classification, on which tidal remnants may naturally impact.
Table \ref{TfracGZ} shows the fractions of LERGs and their morphologically matched controls with tidal features

\renewcommand\arraystretch{1.3}
\begin{table*}
	\centering
	\caption{Fraction of LERGs and control galaxies with tidal features for the full sample and various mass and environment subsets.
	The first column defines the subset of the population analysed.
	The second and third columns show the fractions of galaxies with minor tidal remnants, and the final two columns show the fractions of galaxies with major tidal disruption.
	The top row shows the entire population. 
	The next three rows show stellar mass subsets of the population, with \textit{`low mass'} defined by the lowest quartile of the stellar mass distribution of our LERG sample, \textit{`intermediate mass'}  by the interquartile range, and \textit{`high mass'} by the upper quartile (see panel b of Figure \ref{LERGcontDists}).
	Environmental subsets of the population are shown in the lowest three rows.}
		\begin{tabular}{l c c c c c c}
		\hline
		\textbf{Galaxy subset} & &\textbf{LERG - minor} & \textbf{Control - minor} & & \textbf{LERG - major} & 		\textbf{Control - major}\\
 		& & \textbf{[\%]} & \textbf{[\%]} & & \textbf{[\%]} & \textbf{[\%]}\\
		\hline \hline
		
		\textit{All} & & $21.4_{-1.0}^{+1.1}$ & $22.3_{-0.5}^{+0.5}$ & & $7.2_{-0.6}^{+0.6}$ & $5.0_{-0.2}^{+0.2}$\\
		
		\hline

		\textit{Low mass} & & $15.9_{-1.8}^{+1.8}$ & $14.3_{-0.8}^{+0.8}$ & & $10.0_{-1.5}^{+1.5}$ & $3.2_{-0.4}^{+0.4}$\\

		\textit{Intermediate mass} & & $21.6_{-1.5}^{+1.5}$ & $24.0_{-0.7}^{+0.7}$ & & $6.2_{-0.9}^{+0.9}$ & $4.6_{-0.3}^{+0.3}$\\

		\textit{High mass} & & $26.4_{-2.2}^{+2.0}$ & $27.3_{-1.0}^{+1.0}$ & & $6.5_{-1.2}^{+1.2}$ & $7.8_{-0.6}^{+0.6}$\\
 
		\hline
		
		\textit{Field galaxies} & & $19.8_{-1.5}^{+1.3}$ & $21.3_{-0.6}^{+0.6}$ & & $7.8_{-1.0}^{+1.0}$ & $4.6_{-0.3}^{+0.3}$\\

		\textit{Group galaxies} & & $25.5_{-1.7}^{+1.7}$ & $23.5_{-0.7}^{+0.7}$ & & $6.7_{-1.1}^{+0.9}$ & $5.4_{-0.4}^{+0.4}$\\

		\textit{Cluster galaxies} & & $8.8_{-2.9}^{+2.9}$ & $23.0_{-2.0}^{+2.0}$ & & $5.9_{-2.0}^{+2.0}$ & $6.1_{-1.1}^{+1.1}$\\
 
		\hline

		\end{tabular}
\label{Tfrac}
\end{table*}

\begin{table*}
	\centering
	\caption{Fraction of LERGs and control galaxies with tidal features for the morphologically-matched sample and various mass and environmental subsets.
	The layout is as for Table \ref{Tfrac}.}
		\begin{tabular}{l c c c c c c}
		\hline
		\textbf{Mass bin} & &\textbf{LERG - minor} & \textbf{Control - minor} & & \textbf{LERG - major} & 		\textbf{Control - major}\\
 		& & \textbf{[\%]} & \textbf{[\%]} & & \textbf{[\%]} & \textbf{[\%]}\\
		\hline \hline
		
		\textit{All} & & $17.4_{-1.1}^{+1.1}$ & $20.8_{-0.7}^{+0.7}$ & & $5.3_{-0.7}^{+0.7}$ & $3.4_{-0.3}^{+0.3}$\\
		
		\hline

		\textit{Low mass} & & $12.4_{-2.0}^{+2.0}$ & $10.8_{-1.1}^{+1.1}$ & & $5.6_{-1.6}^{+1.6}$ & $1.4_{-0.4}^{+0.4}$\\

		\textit{Intermediate mass} & & $19.7_{-1.6}^{+1.8}$ & $23.4_{-1.0}^{+1.0}$ & & $5.2_{-0.9}^{+0.9}$ & $3.4_{-0.4}^{+0.4}$\\

		\textit{High mass} & & $17.1_{-2.4}^{+2.4}$ & $24.8_{-1.4}^{+1.5}$ & & $5.3_{-1.6}^{+1.6}$ & $5.5_{-0.7}^{+0.9}$\\
 
		\hline
		
		\textit{Field galaxies} & & $14.5_{-1.6}^{+1.6}$ & $20.3_{-1.0}^{+1.0}$ & & $7.7_{-1.2}^{+1.2}$ & $2.6_{-0.4}^{+0.4}$\\

		\textit{Group galaxies} & & $22.2_{-1.9}^{+1.9}$ & $20.9_{-1.0}^{+1.1}$ & & $3.4_{-0.8}^{+0.8}$ & $3.6_{-0.4}^{+0.4}$\\

		\textit{Cluster galaxies} & & $5.7_{-2.9}^{+2.9}$ & $23.9_{-3.3}^{+3.3}$ & & $1.4_{-1.4}^{+1.4}$ & $8.9_{-2.2}^{+2.2}$\\
 
		\hline
		\end{tabular}
\label{TfracGZ}
\end{table*}


\subsection{Are Tidal Features More Common in Low- or High-Mass LERGs?}
\begin{figure*}
	\centering
	\subfigure[main control]{\includegraphics[width=0.495\textwidth]{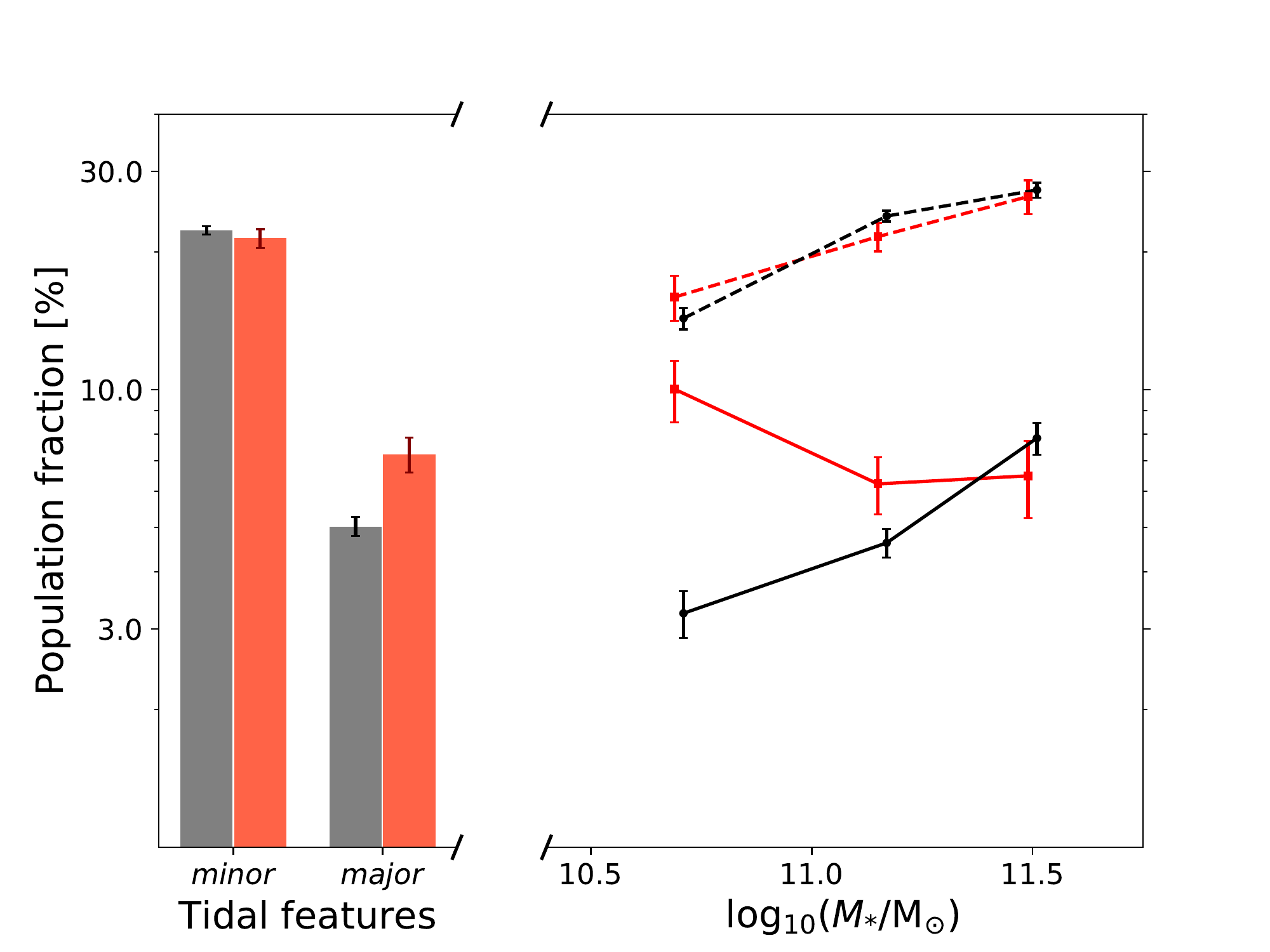}}
	\subfigure[morphologically-matched control]{\includegraphics[width=0.495\textwidth]{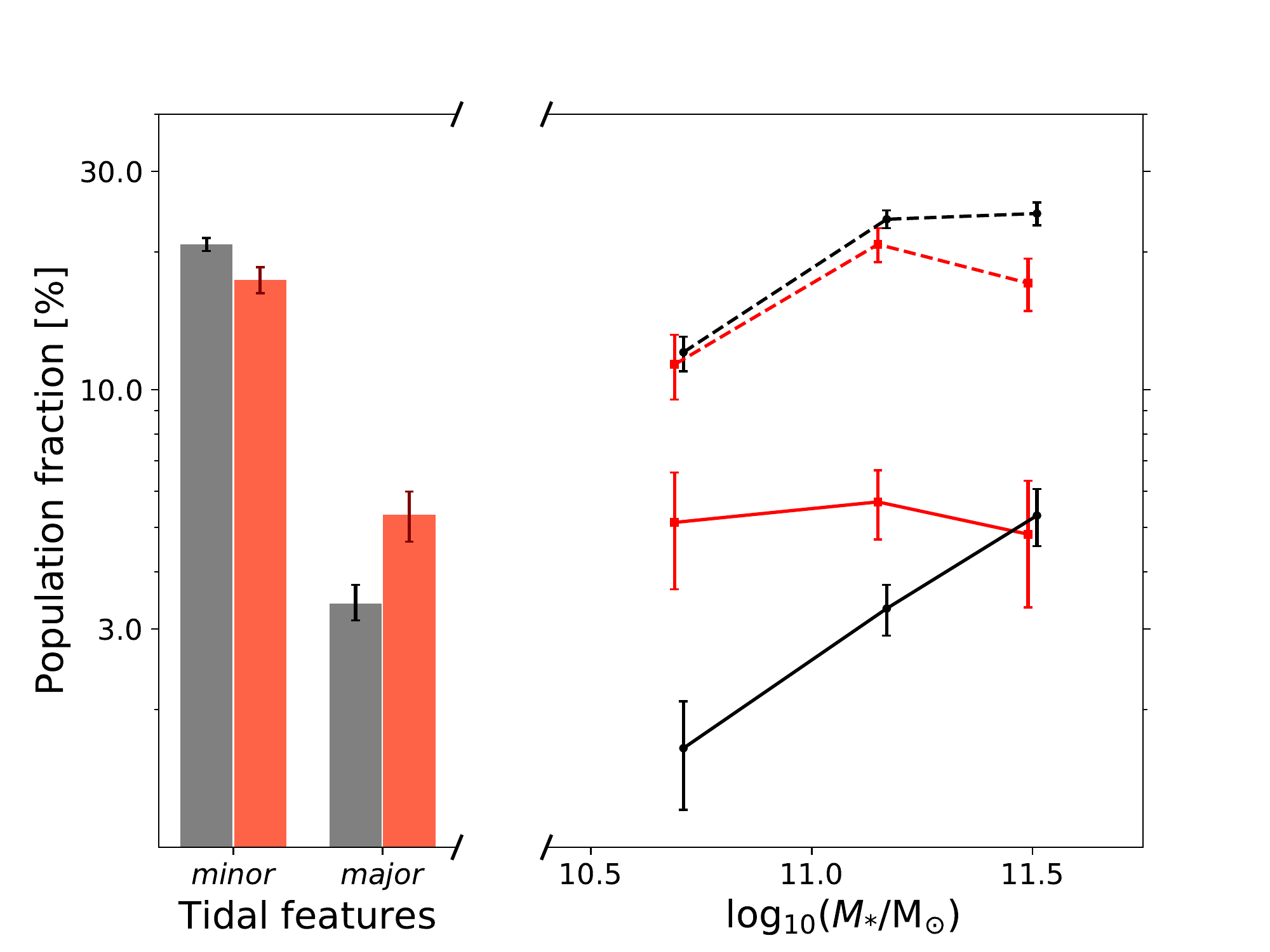}}
\caption{The fractions of LERGs and controls showing minor and major tidal features, with panel a used to show the larger main control sample, and panel b showing the morphologically-matched subset.
The overall fractions are shown in the bar plots on the left, while the fractions in different stellar mass bins, specifically the interquartile range and the two extreme quartiles, are shown on the right.
Here the dashed lines show the minor tidal feature fractions, and the solid lines show the fractions of galaxies with major tidal features.
The LERG and control populations on the right are horizontally offset from each other for clarity.
The excess of LERGs undergoing major interactions is clearly shown to be driven by the low mass end of our sample.
Red is used to represent the LERG population and grey/black represents the control sample.}
\label{majminfracs}
\end{figure*}

To investigate whether the fraction of LERGs having tidal features evolves with stellar mass, we repeat the analysis for the outer quartiles and the interquartile range of the stellar mass distribution of our LERG host galaxies.
Doing this we find that the excess of LERGs having major tidal features is driven by `lower-mass', $\log_{10}(M_{*}/\text{M}_{\odot}) \lesssim 11$, galaxies.
When only the lower quartile of our LERG stellar mass distribution, $\log_{10}(M_{*}/\text{M}_{\odot}) < 10.97$,  is considered, $10.0 \pm 1.5\,$\% of LERGs are observed to have these major remnants compared to $3.2 \pm 0.4\,$\% of control galaxies. 
That is to say, an excess of major tidal features in LERGs at lower stellar masses is seen with $>4\sigma$ confidence.
When the morphologically-matched sample is considered this excess is seen with $>2.5\sigma$ confidence for this mass range, and rising to $>3.5\sigma$ confidence for \text{$\log_{10}(M_{*}/\text{M}_{\odot}) < 11.16$} (i.e. if we stack the two lowest mass quartiles).
For the interquartile range and the upper quartile of the stellar mass range, this excess of major mergers in the LERG population is not observed to a significant level.
This evolution with stellar mass for LERGs to be associated with major tidal features is shown in Figure \ref{majminfracs}, and in Table \ref{Tfrac}.

When compared to the morphologically-matched control subset, LERGs are observed to have a deficit of minor tidal remnants with $>2.5\sigma$ confidence in the highest mass bin, i.e., \text{$\log_{10}(M_{*}/\text{M}_{\odot}) > 11.3$}.
At lower masses, or without controlling for morphology, the fraction of LERGs with minor tidal features is consistent with the control sample.
As with the observations of major tidal features, the fractions of LERGs and control galaxies in different mass bins with minor tidal features is shown in Figure \ref{majminfracs} as well as in Table \ref{TfracGZ}.

\subsection{The Influence of Galaxy Environment on the Likelihood of LERGs to Have Tidal Features}

\begin{figure*}
	\centering
	\subfigure[main control]{\includegraphics[width=0.495\textwidth]{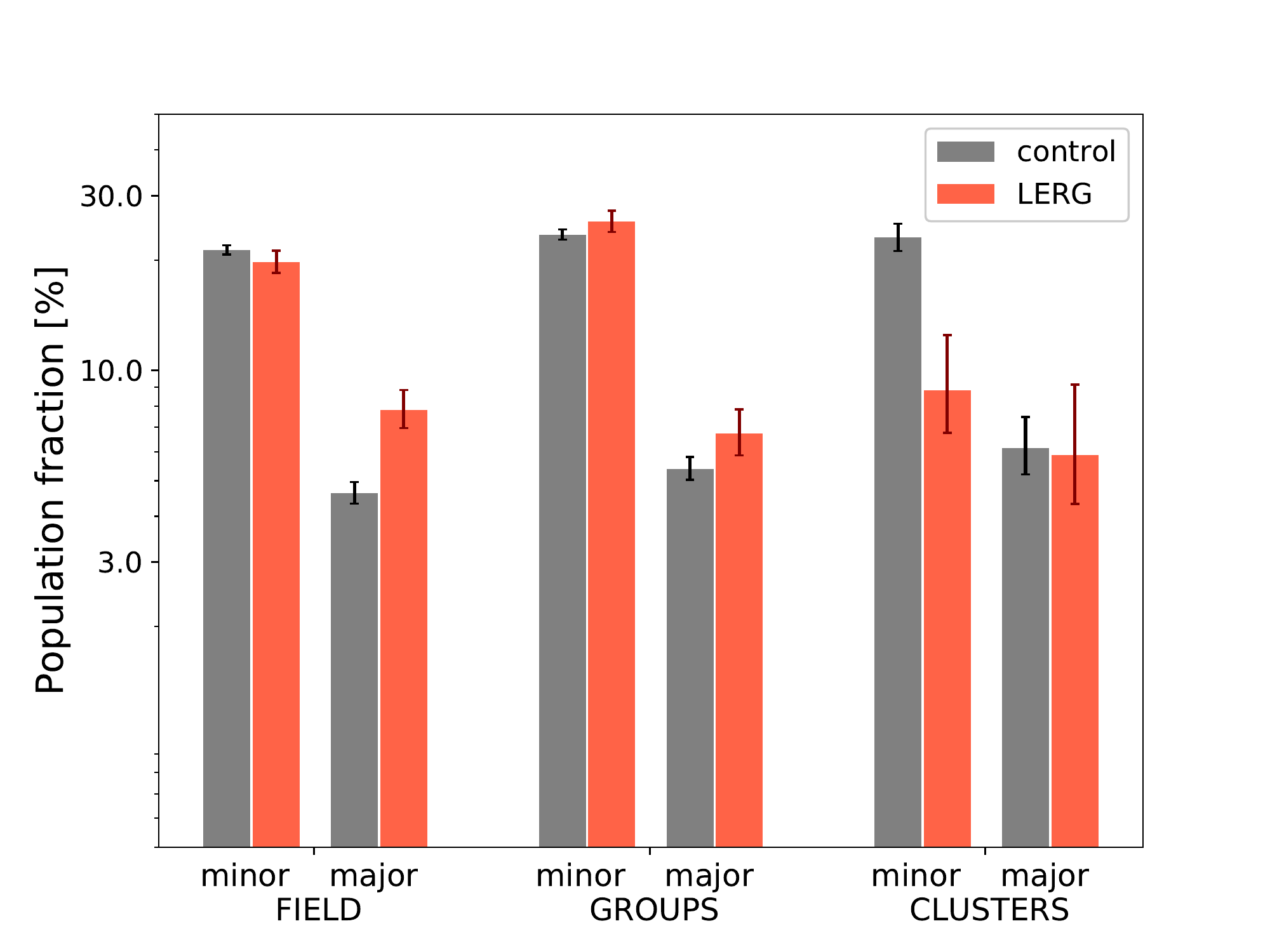}}
	\subfigure[morphologically-matched control]{\includegraphics[width=0.495\textwidth]{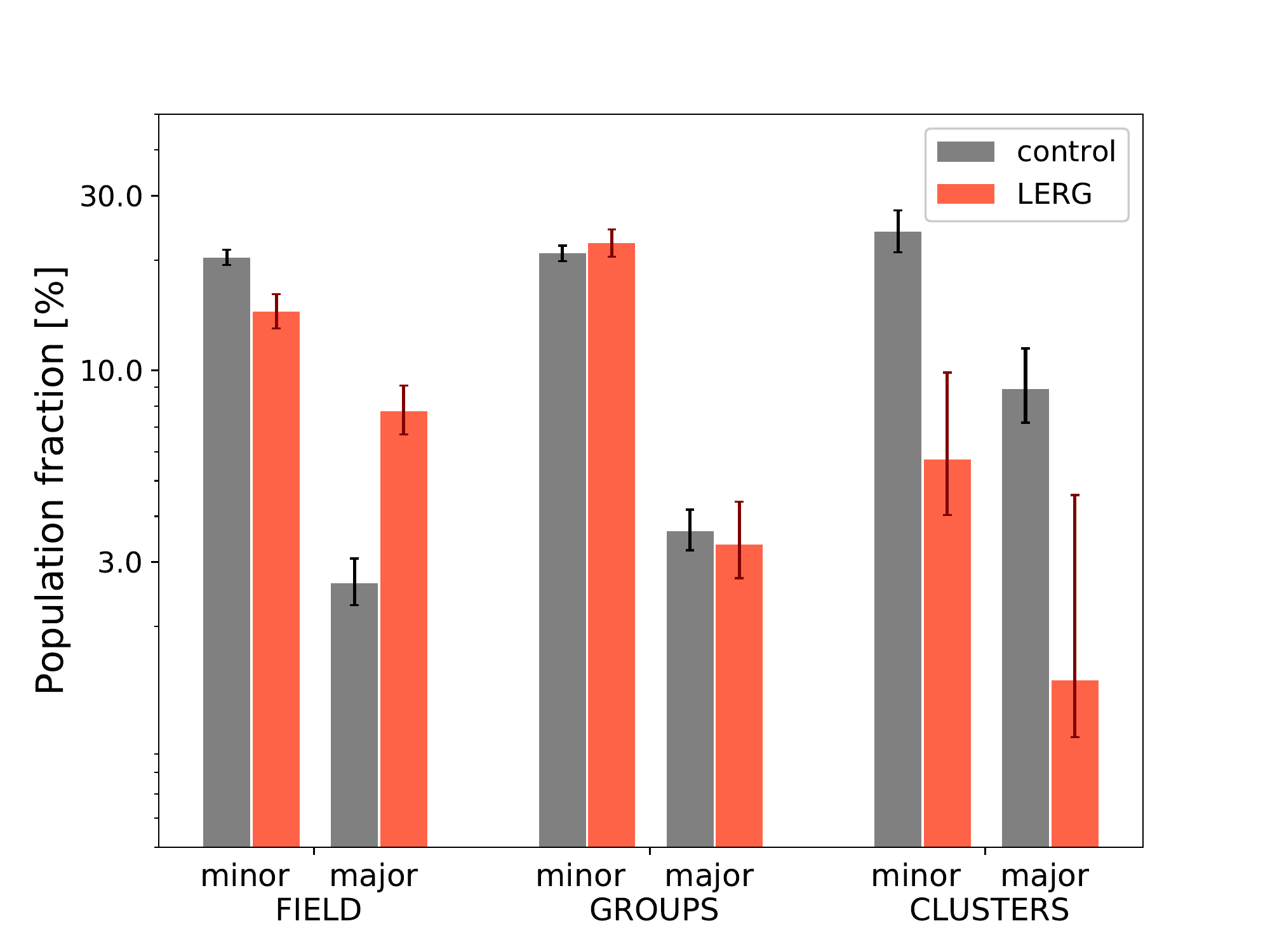}}
\caption{The effect of galaxy environment on the LERG merger fractions with respect to the control population. 
For each environment (field, groups or clusters) the minor merger fractions for the LERGs (red) and controls (grey) are shown on the left, whilst the major merger fractions are shown on the right. 
The left-hand panel shows the full LERG sample and control, whilst the right-hand panel shows the morphologically-matched subset.}
\label{envfracs}
\end{figure*}

We additionally compared the fraction of LERGs and control galaxies with tidal features within different large-scale environments, i.e, whether the galaxy is in the field, a group, or within a cluster.
For the purposes of differentiating between groups and clusters, we segregate these structures at halo masses of $10^{14}\,\text{M}_{\odot}$.
That is to say structures with $M_{180} <10^{14}\,\text{M}_{\odot}$ are classed as groups, and those with $M_{180} > 10^{14}\,\text{M}_{\odot}$ are considered to be clusters \citep{Lofthouse2018, Barsanti2017}.
Galaxies in very low mass groups, $M_{180} < 10^{12.5}\,\text{M}_{\odot}$, are treated as field galaxies \citep{Gordon2018}.
The merger fractions of LERGs and controls in these different environments are presented in Tables \ref{Tfrac} and \ref{TfracGZ}, and in Figure \ref{envfracs}, as well as being described below.

\subsubsection{Field galaxies}
When only field galaxies are considered the excess of major tidal features in LERGs is detected at the $\sim 3\sigma$ level, with $7.8 \pm 1.0 \,$\% of LERGs having such features compared to $4.6\pm0.3$\% of the control sample. 
As with the whole population this excess is greatest in the lower mass regime of our sample. 
To maintain a sizeable population whilst subsetting on multiple parameters, here we simply look at galaxies above and below the median stellar mass of our sample, $\log_{10}(M_{*}/\text{M}_{\odot}) = 11.16$. 
We note that the LERG excess of major tidal features is seen at $ >2.5\sigma$ in the lower mass field population, compared to $< 2\sigma$  for the higher mass galaxies.
Requiring the control population to be matched on morphology does not influence this observation.
Morphology-matching does show that whilst $20.3 \pm 1.0\,$\% of control galaxies have minor tidal features, only $14.5 \pm 1.6\,$\% of LERGs exhibit such morphological disturbances, a $\sim 3\sigma$ deficit. 
This appears to be driven by stellar mass with this excess becoming insignificant at $\log_{10}(M_{*}/\text{M}_{\odot}) < 11.16$.

\subsubsection{Galaxy groups}
In galaxy groups when the whole mass distribution is considered, no significant trends are observed for the fraction of  LERGs with tidal features (either major or minor) to differ from control population.
However, as with the overall analysis, and the analysis on field galaxies, differences in the fractions of control galaxies and LERGs with major morphological disruptions are noticed in different mass regimes.
In the lower quartile of the stellar mass distribution the excess of major morphological disturbances in LERGs is seen at the $\sim 3\sigma$ level, with $14.0 \pm 0.4\,$\% of LERGs having such disruption to their morphology compared to $3.2 \pm 0.9\,$\% of control galaxies.
At the highest stellar masses this trend is inverted.
As few as $2.6 \pm 1.0\,$\% of LERGs with $M_{*} > 10^{11.3}\,\text{M}_{\odot}$ in groups have major morphological disruption in comparison to $7.6 \pm 0.9\,$\% of the control sample, a $>3.5\sigma$ deficit.
The requirement for the control sample to be matched to the LERGs on morphology has no impact in these results.

\subsubsection{Galaxy clusters}
In galaxy clusters the fraction of LERGs with major tidal features is consistent with the control population.
Even at in the lower mass galaxies, where an excess of LERGs with major tidal features appears to be strongest, such an effect is only detected with $< 2\sigma$ confidence.
Furthermore a deficit in LERGs with major tidal features is noted at $>2.5\sigma$ confidence when LERG morphology is controlled for.

Considering the galaxies with minor disturbances to their morphology,  $8.8 \pm 2.9\,$\% of LERGs in clusters display such features.
This presents a $\sim 4 \sigma$ deficit relative to the control population, where $23.0 \pm 2.0\,$\% of galaxies show minor morphological disruption.
No significant variance of this result is observed with changes in the stellar mass range analysed, or with the requirement for LERG morphology to be matched in the control.

%
%

\section{Interpretation of observations and discussion}
\label{discussion}

\subsection{Potential Sources of Bias}
\subsubsection{Spectroscopic targeting and completeness}
Both our LERG and control populations are selected from DR7 of the SDSS spectroscopic catalogue \citep{Abazajian2009}.
One of the limitations of this catalogue is the effect of fibre-collisions that prevent any two spectroscopic fibres on the same plate being positioned closer together than 55" \citep{Strauss2002, Patton2008}.
Consequently, spectroscopic completeness is impeded in regions of the sky with a high target density \citep[e.g.,][]{Yoon2008, Robotham2010, Gordon2017, Gordon2018a}.

Due to fibre-collision induced incompleteness, SDSS targeting algorithms prioritise some targets ahead of others. 
Specifically with respect to this work, objects identified as quasar candidates are prioritised ahead of the main galaxy sample \citep{Strauss2002, Blanton2003}. 
One might expect a low-$z$ sample of LERGs, such as ours, which are predominantly red, extended sources, not to be selected as quasar candidates based on their optical photometry.
However, quasar candidates in SDSS are identified both by their optical photometry, and the presence of radio emission via cross-matching with FIRST \citep{Richards2002}.
Should our LERG sample have a higher targeting priority in dense regions of sky, then this could potentially bias the comparative merger fractions we observe in the LERGs and control galaxies.

The entirety of our LERG sample and more than $99\,$\% of our control population are included in the SDSS DR7 main galaxy sample.
Consequently, reanalysing our observations limited to just SDSS main sample galaxies has no impact on our results.
As an additional check to ensure there is no difference in the density of potential spectroscopic targets close proximity to our LERG and control samples, we analyse the number of neighbours observed in the SDSS DR7 photometric catalogue, and with $14.5<r<17.77$, around both populations.
We find that the number of neighbours within 55" is consistent for the LERG and control populations, with an average of $1.18 \pm 0.06$ and $1.10 \pm 0.03$ neighbours respectively.
Reducing the search radius to 25", i.e. less than half the fibre-collision limit, we observe an average of $0.33 \pm 0.06$ neighbours per LERG, and $0.29 \pm 0.03$ neighbours per control galaxy.
We are thus confident that our merger fractions are not biased due to differences in on-sky target densities.

\subsubsection{Treating low mass halos as the field}
In matching the LERG and control samples on large-scale environment, we chose to consider galaxies within halos with $M_{180} < 10^{12.5}\,\text{M}_{\odot}$ as being in the field rather than in groups.
However, given that LERGs are frequently observed in over-dense environments, it may be the case that LERGs classified in the field may be more likely than their control galaxies to be a member of a low mass group. 
If this were the case, it could explain the increased fraction of our `field' LERGs that exhibit tidal features relative to the control sample.

\begin{figure}
	\centering
	\includegraphics[width=\columnwidth]{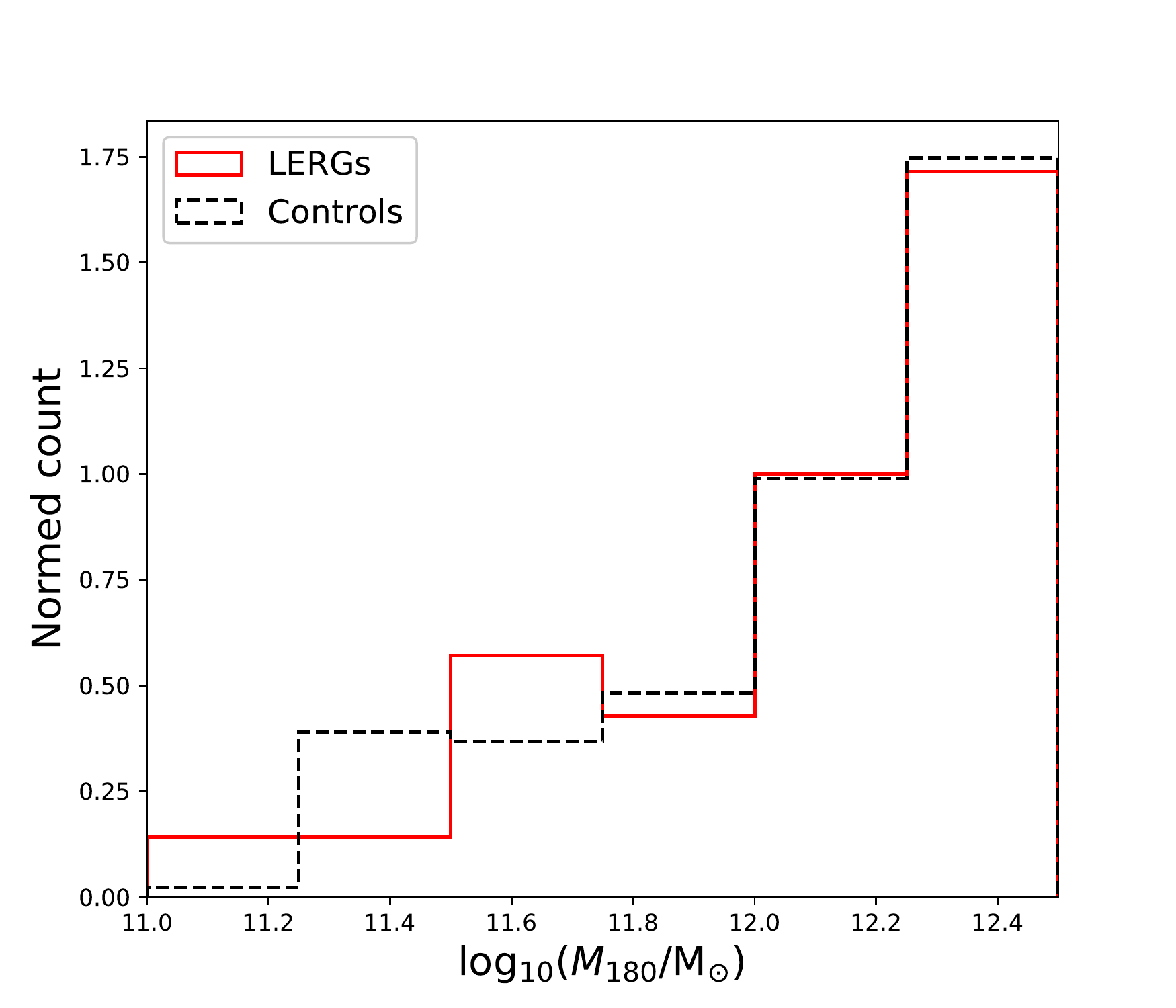}
	\caption{The distributions of halo mass estimates for galaxies assigned to groups with $M_{180}<10^{12.5}\,\text{M}_{\odot}$, and thus treated as field galaxies in our analysis.
	The red solid line shows the low mass groups containing LERGs, and the black dashed line the groups containing their control galaxies.
	Applying a two-sample KS test to these distributions returns $p = 0.98$, consistent with both distributions being drawn from the same parent population.
	Halo mass estimates obtained from the \citet{Yang2007} SDSS group catalogue.}
\label{lowmasshalos}
\end{figure}

In our LERG and control populations classified as being field galaxies, $18.9_{-2.8}^{+3.6}\,$\% and $19.9_{-1.3}^{+1.4}\,$\% respectively reside in these low mass halos.
Additionally, for the galaxies in the low mass groups that we have classified as the field, the tendency for LERGs to be found in over-dense environments may translate to them being in higher mass halos than the control galaxies.
Comparing the halo mass distributions of these low mass groups shows no difference between the groups hosting LERGs and control galaxies, with a KS derived $p$-value of 0.98 (see also Figure \ref{lowmasshalos}).
Furthermore, we re-conduct our analysis of the field galaxy population using only those galaxies hosted by low mass groups.
Here we find no excess of LERGs showing major morphological disturbances, with $2.8 \pm 1.4\,$\% of LERGs displaying such features compared to $3.5_{-1.1}^{+0.9}\,$\% of their controls.
This demonstrates that the observation of excess major tidal disruption in field LERGs is not driven by LERGs residing in the lowest mass groups.

\subsubsection{Weighting of the control sample}
\label{ControlWeights}
In conducting this work each control galaxy was given an equal weighting, and, in order to maximise the sample size, all possible control galaxies were used for any particular analysis.
As stated in Section \ref{data}, where possible we have selected six control galaxies per LERG.
However, in $\sim10\,$\% of cases fewer than six control galaxies could be found for a LERG, and in these cases as many controls as can be found that satisfy the matching criteria are included.
Thus, there is the potential for this variance in the number of control galaxies available per LERG to influence our results.
Given that just $\sim10\,$\% of LERGs have fewer than six control galaxies selected, then a quick test of such an effect would be to restrict our analysis to those LERGs with the maximum number of available control galaxies (252 LERGs and 1512 control galaxies).
Doing this, we find no substantial changes to our results, with all of our statistically significant observations remaining above $3\sigma$ confidence. 
Our observations are thus not significantly affected by the limited number of control galaxies available to some of the selected LERGs.

\subsubsection{Interpreting Merger Scale From Tidal Feature Intensity}

Although we have classified galaxy images based on the intensity of their tidal features (should any be present), does this necessarily translate to merger intensity?
In particular, in the very latest stages of major merger, even the most substantial of tidal features will dissipate over time as the two galaxies coalesce.
Consequently a very late stage major merger may show only minor disruptions to the extended LSB morphology \citep[for example see Figure 1. in][]{Lotz2008}.

The degeneracy in the origin of minor LSB features naturally complicates attributing them to either a minor or major merger.
It is anticipated that straight tidal streams (e.g. as shown in column 2, row 2 of Figure \ref{egClass}) are the result of recent minor mergers rather than being older remnants \citep{Duc2015}.
Furthermore, the shell structures seen around many early type galaxies (e.g. column 2, row 3 of Figure \ref{egClass}), and which we have classified as a minor morphological disruption, are frequently associated with low-intermediate mass ratio mergers \citep[i.e with mass ratios in the range 1:3 to 1:10,][]{Kaviraj2010, Duc2015, Pop2018, Kado-Fong2018}.
Considering these points in combination with the relatively short period during which a major merger exhibits only minor tidal features, any contamination of the minor disturbance sample by late stage major mergers should be minimal.
For these reasons we adopt the approach that minor tidal features are the result of minor mergers, whereas major features are due to major mergers, typical of previous works involving classification of morphological disruptions in large samples of galaxies \citep[e.g.][]{Kaviraj2014, Kaviraj2014a, Morales2018}.

\subsection{The Influence of Major Mergers on LERG Evolution}
\subsubsection{LERGs can evolve from a major merger}
Our observations show a $>3\sigma$ excess of LERGs undergoing major mergers in comparison to a control sample, and this excess is shown to be strongest ($>4\sigma$ confidence) at lower stellar masses.
This mass trend of the LERG major merger fraction persists in all environments. 
Although only seen at less than $2\sigma$ confidence for galaxies within clusters, even here it is at the lower end of the stellar mass distribution of our sample where any potential excess of major mergers in LERGs is noted.
This demonstrates that major mergers can result in the production of a LERG, particularly in field and group galaxies, and at stellar masses below $\sim 10^{11}\,\text{M}_{\odot}$. 

Prior observations have shown mixed results with regard to the likelihood of major mergers being involved in the evolution of a galaxy into a LERG.
For galaxies with $10 < \log_{10}(M_{*}/\text{M}_{\odot}) < 12$, \citet{Sabater2013} demonstrated that LERGs were more likely than non-LERGs to be involved in one-on-one interactions with other galaxies.
On the other hand, \citet{Ellison2015} demonstrated that LERGs were no more likely to be in a close galaxy pair than other similar galaxies after controlling for galaxy properties and large scale structures, indicating that major mergers are not the primary trigger mechanism for LERGs.
In post-merger galaxies showing tidal remnants however, \citet{Ellison2015} do observe a slight, but insignificant, excess of LERGs relative to their control sample, potentially consistent with our results and those of \citet{Sabater2013}.
Furthermore, while \citet{Ellison2015} do not state the stellar mass range of their sample, their Figure 1 demonstrates that their LERGs in pairs are drawn from a population dominated by galaxies with $M_{*}>10^{11}\,\text{M}_{\odot}$.
It is thus possible that their work either does not include the low-mass galaxies that are driving the excess of major mergers in LERGs that we observe, or that these constitute a less substantial fraction of their LERG sample.

Whilst our observations show that major mergers clearly play a role in the triggering of some low-mass LERGs, even in this mass regime only $\sim10\,$\% of LERGs are currently experiencing such large-scale interactions. 
Major mergers therefore are not the dominant pathway to LERG activity.
Hence, for the remaining \text{$\sim 90\,$\%} of the low-mass LERG population, and for LERGs of higher stellar masses, other trigger mechanisms must be invoked.

\subsubsection{Are major merger-driven low-mass LERGs the progenitors of HERGs?}
The relatively low accretion rates associated with the nuclear activity in LERGs suggests that while major mergers can trigger low-mass LERGs, they don't directly fuel the black hole accretion within these galaxies.
Rather the fuel source may be internal in origin, and that the accretion on to the supermassive black hole is induced by disk instabilities that are the result of pre- and in-merger gravitational perturbations \citep{Bournaud2011, Nealon2015, Gatti2016, King2018}.
Such a mechanism has the potential to provide a more limited fuel supply than one might expect if the fuel originated from the donor component of the merging system, and hence explain the radiatively inefficient accretion mode observed.
Moreover, radiatively inefficient black hole accretion may be just the first step of nuclear activity in these galaxies.
It is reasonable to expect that even for efficiently accreting AGN there exists a short phase where the accretion rate is radiatively inefficient \citep{Sabater2013}.
In this scenario, and if the merger is gas-rich, these merger-induced LERGs could then evolve into HERGs once the gas from the merger has fallen into the central engine providing the opportunity for more radiatively efficient accretion to take place.
Such a process would be consistent with observations showing that Seyfert-like emission line ratios peak towards the end of the merger process \citep{Carpineti2012}, and the association between mergers and powerful radio galaxies seen at high-$z$ \citep{Chiaberge2015}.

These major merger-driven LERGs only represent a small fraction of the LERG population, and thus may not be typical of inefficiently accreting RLAGN. 
The excess of major mergers in LERGs is seen most strongly at $\log_{10}(M_{*}/\text{M}_{\odot}) < 10.97$.
It is interesting to note that while the median of our LERG stellar mass distribution is $\sim 0.2\,$dex higher than this,  the median stellar mass of HERGs in the \citet{Best2012} catalogue with $z<0.07$ is $\log_{10}(M_{*}/\text{M}_{\odot}) \sim 10.9$, a difference of $<0.1\,$dex.
In other words, the masses of LERGs where we observe the strongest excess of major mergers, are more typical of the HERG population than of the broader LERG population, and, if merger-driven LERGs are potential progenitors of HERGs, one would expect their masses to be broadly consistent.

Should this description be an accurate representation of the physics at play, then low-mass LERGs associated with major merger systems represent young AGN, ostensibly at the point of trigger.
These objects may thus be associated with relatively compact radio morphologies compared to LERGs that are not merger-driven, due to the jet having limited propagation time.
Whilst we have made no attempt to do so in this work, a comparative analysis of the radio properties of merging versus non-merging LERGs presents a compelling opportunity for follow up.
The next generation suite of high-resolution radio surveys such as the Very Large Array Sky Survey\footnote{\url{https://science.nrao.edu/science/surveys/vlass}} (VLASS, Lacy et al. in prep) should be well suited to such an analysis.

Such a smooth transition from radiatively inefficient to efficient accretion modes as the AGN evolves may of course be an oversimplification.
It has been demonstrated that AGN can `flicker' on and off over timescales $\lesssim 10^{5}\,$ years \citep{Schawinski2015, King2015, Comerford2017}.
In radio quiet AGN, a drop in accretion rate below $\sim 0.01\, \dot M_\text{Edd}$ would appear as an AGN being switched off.
In RLAGN however, the presence of the relativistic jet ensures the galaxy is still detected as an AGN even at low accretion rates.
This presents an alternative possibility to a steady LERG-to-HERG evolution, in that some LERGs may simply be the low-accretion phase of RLAGN flickering.
However, in general, there are differences in both the stellar populations and radio properties of LERGs and HERGs \citep[e.g.][]{Baum1992, Buttiglione2010, Best2012}.
Consequently a comparison of these properties in major merger-driven LERGs and HERGs of similar mass will be required to test such a hypothesis, and is beyond the scope of this work.

\subsection{The Role of Minor Mergers in the Evolution of LERGs}
\subsubsection{LERGs are not primarily fuelled by minor mergers}
Given the requirement for a low Eddington scaled accretion rate in LERGs, minor mergers have been proposed as a potential pathway with which to introduce a limited fuel supply to the central engine \citep{Kaviraj2014, Pace2014, Ellison2015}.
Should this be the case, then LERGs are expected to have a higher fraction of tidal features associated with minor mergers than a control population \citep{Kaviraj2014, Kaviraj2014a}, and testing this hypothesis is one of the principle aims of this work.
We observe no such excess of minor mergers in our LERG sample, suggesting that these events do not play a substantial role in the fuelling of LERGs.

The validity of our test is dependent on a couple of assumptions regarding the visibility of the remnants of minor mergers.
The merging system must be able to leave a detectable remnant with a surface brightness of $\mu_r < 28\,\text{mag}\,\text{arcsec}^{-2}$.
\citet{Ji2014} investigated the visibility of tidal remnants resulting from different scale mergers involving a simulated galaxy with $\log_{10}(M_{*}/\text{M}_{\odot}) \sim 10.4$\footnote{This value is determined from the values of bulge stellar mass and disk stellar mass provided in Table 1 of \citet{Ji2014}.}. 
Their work demonstrates that such a galaxy experiencing a $>$1:10 merger can produce tidal features with $\mu_{r} < 28\,\text{mag}\,\text{arcsec}^{-1}$.
As the majority of satellites to massive galaxies are dwarfs with $M_{*} < 10^{9}\,\text{M}_{\odot}$  \citep[e.g.,][]{Loveday1997, DeRijcke2006}, then clearly a substantial number of very-minor mergers will go undetected.
This may be particularly true for RLAGN given their increased number of satellites relative to inactive galaxies \citep{Pace2014}.
Furthermore, if the minor merger is to directly fuel the LERG, then the merger remnant must remain visible long enough for the gas from the donor galaxy to fall in to the central engine of the recipient.
Based on the stellar population ages of AGN, this process is estimated to take several hundred Myr \citep{Tadhunter2005, Bessiere2014, Shabala2017}.
At surface brightnesses of $\mu_r < 28\,\text{mag}\,\text{arcsec}^{-2}$, \citet{Ji2014} show that the remnant from a 1:6 mass ratio merger should be observable for $>2\,$Gyr post-merger.
The visibility timescale of the merger remnant should therefore not prohibit the association of low to moderate mass ratio mergers with nuclear activity.
Consequently we can make the statement that moderate to minor mergers are not the primary fuel supply for LERGs.

\subsubsection{Minor mergers inhibit LERG activity in clusters}
Although minor mergers do not preferentially trigger LERGs, it may not be accurate to say these events play no role in the evolution of an inactive galaxy into a LERG.
We observe a significant ($4\sigma$) deficit of minor mergers in LERGs residing within clusters.
Beyond just failing to contribute to the triggering of a LERG, these observations indicate that minor mergers may actually prevent galaxies within clusters from evolving in to LERGs.

Within clusters it has been shown that different regions of the cluster environment provide different opportunities for AGN fuelling \citep[e.g.][]{Haines2012, Pimbblet2013, Gordon2018}.
The cluster core for instance may allow for cooling flows onto a galaxy, a widely hypothesised mechanism for LERG fuelling \citep[e.g.][]{Gaspari2013, Gaspari2017, Tremblay2016}.
The outer regions of clusters provide more opportunities for low-speed interactions such as mergers, and infalling galaxies may experience ram-pressure stripping. 
Both of these mechanisms are known to be associated with nuclear activity in galaxies \citep[e.g.][]{Sanders1988, Poggianti2017}.
Ergo, whilst we control for cluster membership, not controlling for position within the cluster may bias this analysis, i.e. if a central LERG is compared to a satellite control galaxy.

Using the brightest cluster galaxy (BCG) flag from the \citet{Yang2007} catalogue, we can crudely segregate galaxies into centrals and satellites.
Doing this the galaxies in our sample selected as satellites all have projected radii from the BCG of $>0.5R_{180}$, with $85\,$\% of these at $R > R_{180}$.
When limiting the control sample to just those with the same central/satellite classification as the LERGs, the deficit of minor merger in LERGs is still observed at $>3.5\sigma$ significance.
This is dominated by the satellite galaxy population, where the of LERG minor merger deficit is seen at the $3\sigma$ level.
On the other hand, in BCGs the minor merger fractions of LERGs and control galaxies are consistent, with $25.0 \pm 12.5\,$\% of BCG LERGs and $30.3 \pm 9.1\,$\% of BCG controls observed to be experiencing such interactions.

This suggests that minor mergers in the cluster core do not inhibit LERG activity.
However, we note that just twelve galaxies in our sample, five of which are LERGs, are BCGs. 
Indeed only 18 LERGs in total within our sample lie within halos of $M_{180}>10^{14}\,\text{M}_{\odot}$
That is to say, $28\pm11\,$\% of our LERGs in clusters are BCGs.
Given the frequent association of LERGs with BCGs \citep[e.g.][]{Tremblay2018}, this might seem to be a relatively low number.
To check that such a low BCG fraction amongst our cluster LERG sample should be expected, we cross match the entire \citet{Best2012} catalogue of LERGs at $z<0.07$ (605 galaxies) with the \citet{Yang2007} group catalogue.
Of the 42 LERGs found to be in clusters, $33\pm7\,$\% are flagged as BCGs, consistent with what we observe in our data.
Repeating this test with $z<0.2$ shows $\sim 50\,$\% of cluster LERGs to be BCGs, indicating that the low fraction of our cluster LERGs that are BCGs may be an effect of the low-redshift, $z<0.07$, nature of this work.
Indeed, \citet{Ching2017} demonstrate that low-power, $L_{1.4\,\text{GHz}} < 10^{24}\,\text{W}\,\text{Hz}^{-1}$, radio AGN are not significantly more likely to be central galaxies than their radio quiet counterparts once matched on stellar mass and colour.
A consequence of the limited depth of our sample is that the median radio luminosity of our LERGs is an order of magnitude lower than this (see Figure \ref{rpowerdist}).
Therefore one might not expect our sample of LERGs to be hosted predominantly by BCGs.
Furthermore, no account is taken of the structure of the clusters, in terms of whether it is relaxed, or actively coalescing with another structure \citep[e.g., as is the case for Abell 1882, see][]{Owers2013}.
Consequently we would urge further studies, making use of even deeper imaging, possibly from future facilities such as, e.g., the Large Synoptic Survey Telescope \citep{LSST2009}, to obtain a larger volume of cluster LERGs with which to analyse merger trends.

\section{Summary}
\label{sum}
In this work we have exploited deep optical imaging to test the role of mergers in the evolution of inefficiently accreting RLAGN.
This was achieved by comparing the prevalence and intensity of tidal features in LERGs ($282$ galaxies) and a control sample ($1,622$ galaxies) matched on redshift, galactic stellar mass and environment.
In particular the depth of imaging, $\mu_{r} < 28\,\text{mag}\,\text{arcsec}^{-2}$, allowed for a large scale analysis of the role of minor mergers, a hypothesised LERG trigger \citep[e.g.,][]{Kaviraj2014, Ellison2015}, in LERG fuelling for the first time.
Our main observations are:

\begin{enumerate}
	
	\item No excess of minor mergers is observed in the LERG population relative to the control in any mass regime or large-scale environment.
	This is at odds with the hypothesis that minor mergers may present a fuel supply with which to power a weakly accreting RLAGN.
	
	\item LERGs in clusters have a minor merger fraction of $8.8 \pm 2.9\,$\% in contrast to $23.0 \pm 2.0\,$\% of control galaxies, a $>4\sigma$ deficit.
	This observation is not only inconsistent with the hypothesis that such events are a major contributor to LERG activity, but also suggests that minor mergers in the cluster environment act to prevent the evolution of an inactive galaxy in to a LERG. 
	
	\item A significant, $>4\sigma$, excess of major mergers is observed in relatively low-mass LERGs. 
	At \text{$M_{*}\lesssim 10^{11}\,\text{M}_{\odot}$}, $10 \pm 1.5\,$\% of these AGN are experiencing such large-scale interactions compared to $3.2 \pm 0.4\,$\% of the control population.
	At higher masses the LERG major merger fraction tends towards that of the control population, with no LERG excess observed at $M_{*} > 10^{11.3}\,\text{M}_{\odot}$. 
	This effect is seen most strongly for field galaxies, but we note that in all environments, any excess of LERG major mergers is seen with the highest confidence at lower stellar masses.
	
\end{enumerate}

In conclusion, our observations show that minor mergers do not fuel LERGs, and are in agreement with an overall picture where the majority of traditional LERGs are powered by the accretion of matter from the halo.
A minority of lower mass LERGs are clearly associated with major mergers. 
In these cases we hypothesis that we may be witnessing a relatively brief phase of low-excitation in a galaxy that may evolve into a HERG.

\acknowledgments

The authors would like to thank the anonymous referee for their highly constructive comments.
YAG, CPO and SAB are supported by NSERC, the Natural Sciences and Engineering Research Council of Canada.
YAG additionally acknowledges the financial support provided by a University of Hull PhD studentship.
KAP acknowledges the support of Science and Technology Facilities Council (STFC), through the University of Hull's Consolidated Grant ST/R000840/1.
MSO acknowledges the funding support from the Australian Research Council through a Future Fellowship (FT140100255).

This publication uses data generated via the \href{https://www.zooniverse.org/}{Zooniverse.org} platform, development of which is funded by generous support, including a Global Impact Award from Google, and by a grant from the Alfred P. Sloan Foundation. Additionally, this research made use of Astropy, a community-developed core Python package for Astronomy \citep{AstropyCollaboration2013, Astropy2018}.

Funding for the SDSS and SDSS-II has been provided by the Alfred P. Sloan Foundation, the Participating Institutions, the National Science Foundation, the U.S. Department of Energy, the National Aeronautics and Space Administration, the Japanese Monbukagakusho, the Max Planck Society, and the Higher Education Funding Council for England. The SDSS Web Site is \url{http://www.sdss.org/}.

The SDSS is managed by the Astrophysical Research Consortium for the Participating Institutions. The Participating Institutions are the American Museum of Natural History, Astrophysical Institute Potsdam, University of Basel, University of Cambridge, Case ßWestern Reserve University, University of Chicago, Drexel University, Fermilab, the Institute for Advanced Study, the Japan Participation Group, Johns Hopkins University, the Joint Institute for Nuclear Astrophysics, the Kavli Institute for Particle Astrophysics and Cosmology, the Korean Scientist Group, the Chinese Academy of Sciences (LAMOST), Los Alamos National Laboratory, the Max-Planck-Institute for Astronomy (MPIA), the Max-Planck-Institute for Astrophysics (MPA), New Mexico State University, Ohio State University, University of Pittsburgh, University of Portsmouth, Princeton University, the United States Naval Observatory, and the University of Washington.

This project used data obtained with the Dark Energy Camera (DECam), which was constructed by the Dark Energy Survey (DES) collaboration. Funding for the DES Projects has been provided by the U.S. Department of Energy, the U.S. National Science Foundation, the Ministry of Science and Education of Spain, the Science and Technology Facilities Council of the United Kingdom, the Higher Education Funding Council for England, the National Center for Supercomputing Applications at the University of Illinois at Urbana-Champaign, the Kavli Institute of Cosmological Physics at the University of Chicago, Center for Cosmology and Astro-Particle Physics at the Ohio State University, the Mitchell Institute for Fundamental Physics and Astronomy at Texas A\&M University, Financiadora de Estudos e Projetos, Fundacao Carlos Chagas Filho de Amparo, Financiadora de Estudos e Projetos, Fundacao Carlos Chagas Filho de Amparo a Pesquisa do Estado do Rio de Janeiro, Conselho Nacional de Desenvolvimento Cientifico e Tecnologico and the Ministerio da Ciencia, Tecnologia e Inovacao, the Deutsche Forschungsgemeinschaft and the Collaborating Institutions in the Dark Energy Survey. The Collaborating Institutions are Argonne National Laboratory, the University of California at Santa Cruz, the University of Cambridge, Centro de Investigaciones Energeticas, Medioambientales y Tecnologicas-Madrid, the University of Chicago, University College London, the DES-Brazil Consortium, the University of Edinburgh, the Eidgenossische Technische Hochschule (ETH) Zurich, Fermi National Accelerator Laboratory, the University of Illinois at Urbana-Champaign, the Institut de Ciencies de l'Espai (IEEC/CSIC), the Institut de Fisica d'Altes Energies, Lawrence Berkeley National Laboratory, the Ludwig-Maximilians Universitat Munchen and the associated Excellence Cluster Universe, the University of Michigan, the National Optical Astronomy Observatory, the University of Nottingham, the Ohio State University, the University of Pennsylvania, the University of Portsmouth, SLAC National Accelerator Laboratory, Stanford University, the University of Sussex, and Texas A\&M University.




\bibliographystyle{aasjournal}
\bibliography{$HOME/Documents/science/Papers/library} 

\end{document}